\documentclass{amsart}
\usepackage{amscd,amsfonts,amssymb,madison}
\begin{document}
\title[Orbifold limits of $K3$: CFT versus Geometry]{Orbifold 
Constructions of $K3$:\\ A Link between\\ Conformal Field Theory and Geometry}
%
\author{Katrin Wendland}
\address{\hspace*{-1.5em}K. Wendland\\
Dept. of Physics and Astronomy\\
UNC at Chapel Hill\\
141 Phillips Hall, CB \#3255\\
Chapel Hill, N.C. 27599}
\email{wendland@physics.unc.edu}
%
\begin{abstract}
We discuss geometric aspects of
orbifold conformal field theories in the moduli space of $N=(4,4)$
superconformal field theories with central charge $c=6$.
Part of this note consists of a summary of our earlier results
on the location of these theories within the moduli space \cite{nawe00,we00} 
and the action of a specific version of mirror
symmetry on them \cite{nawe01}. 
We argue that these results allow for a
direct translation from geometric to conformal field theoretic data. 
Additionally, this work contains a 
detailed discussion of an example which allows the application of various
versions of mirror symmetry on $K3$. We show that all of them agree in
that point of the moduli space.
\end{abstract}
\maketitle
\section*{Introduction}\label{intro}
This note is intended to make a contribution to the understanding of
links between algebraic geometry and theoretical physics, with an emphasis
on geometric aspects of conformal field theory.

From the set up of string theory, a connection to geometry is more or less
obvious, but in general it seems to be hard to formulate it in precise
mathematical terms. Nevertheless, many aspects of string theory have (had)
a strong influence on mathematics, among them 
\textsc{orbifold string theory} and \textsc{mirror symmetry}, 
both of which are  leitmotifs for the present
note. String theory at small coupling is described by a superconformal
field theory (SCFT)  on the world sheet. Therefore, a connection 
between geometry
and SCFT is expected, too. Since SCFTs are well-defined mathematical
objects in their own right, an investigation of direct links to geometry offers
a mathematically safe basis which we are using in this work. 

We  focus on a special type of SCFTs which
is simple enough to carry out a  sound analysis but also 
provides enough non-trivial structure to find interesting links
to geometry:
We investigate
aspects of the moduli space of those SCFTs 
with central charge $c=6$
whose Hilbert space is a representation of a specific $N=(4,4)$ superconformal
algebra $\AA$. Namely, $\AA$ contains an affine
$su(2)_l\oplus su(2)_r$ Kac-Moody algebra at level $1$ \cite{aetal76},
such that all left and right charges 
with respect to a Cartan subalgebra of
$su(2)_l\oplus su(2)_r$ (i.e.\ all doubled spins) are integral. 
We are working with a partial completion $\MM$ of those two components of this
moduli space that are relevant in string theory,  
which for simplicity we call the \textsc{moduli space of
$N=(4,4)$ SCFTs with central charge $c=6$}.

One reason to make such concessions
is the fact that the  space $\MM$ is known explicitly
\cite{na86,se88,ce91,asmo94,nawe00}, and its very
description already allows to draw  links
between geometric and superconformal field theoretic data:
Its two connected
components $\MM^{tori}$, $\MM^{K3}$ are naturally interpreted
as extensions of the geometric moduli spaces of Einstein metrics on
a complex two torus and a $K3$ surface, respectively.
On the other hand, statements about orbifold CFTs and mirror symmetry for
theories in $\MM$ are non-trivial already. 

We will begin with a summary of what is known about the moduli space 
$\MM$. It in particular includes the precise description of the 
location of orbifold CFTs of toroidal theories within $\MM$ that was obtained
in \cite{nawe00,we00}. This requires the determination of certain B-field
values which at first sight might appear to be a physicist's invention
that is hard to assign an intrinsic geometric meaning to. However, 
we will argue that orbifold CFTs on $K3$ allow for an explicit geometric
explanation of the B-field by the use of the classical McKay correspondence.
This viewpoint is somewhat complementary to the one taken in 
\cite{sh01a,sh01b,sh01c} but proves to be
particularly useful in the present context.
A second topic of this note is the discussion of mirror symmetry on 
$\Z_N$ orbifold CFTs on $K3$, which summarizes the results of \cite{nawe01}. 
We use a geometric approach by fiberwise
T-duality on a torus fibration of the manifold under discussion, which
goes back to ideas by Vafa and Witten \cite{vawi95}. It allows us to 
determine the mirror map for $\Z_N$ orbifold CFTs on $K3$ explicitly. 
Since both the geometric and the conformal field theoretic approaches are 
totally under control, it also provides
an explicit translation  between geometric and conformal field theoretic data. 
The third and last part of the present work is devoted to a detailed 
discussion of mirror symmetry for a particular SCFT in $\MM$. This theory
allows a comparison of our approach to mirror symmetry with two other
versions that have been successfully applied to $K3$ before 
\cite{grpl90,ba94,asmo94,do96}. We show that with an emendation due to Rohsiepe
\cite{ro01a,ro01b}
all these approaches are compatible
for our example.

This work is organized as follows:
In Sect.\ \ref{modulispace} we give a brief description of the moduli
space $\MM$ as algebraic space. Section \ref{orbifolds} explains the location
of orbifold CFTs of toroidal theories within $\MM$ and in particular
provides a solution to the ``B-field problem'' in terms of geometric 
structures. In Sect.\ \ref{mirror} we give a somewhat superficial introduction
to those aspects of mirror symmetry that are relevant for our discussion
of $\MM$. The version of mirror symmetry which is induced by fiberwise
T-duality on a specific elliptic fibration of a $\Z_N$ orbifold limit of $K3$
gives another direct link between geometry and SCFT. This is explained
in Sect.\ \ref{cftvsgeo}. Section \ref{example} contains the discussion of
a particular theory in $\MM$. We compare three versions of mirror symmetry
on $K3$ and show that they agree for this theory. We end with a summary
and discussion in Sect.\ \ref{conc}.

The aim of this note is to give a digestible overview on the subject and to
explain the above mentioned example. In particular, 
proofs that are already written up
elsewhere (see \cite{nawe00,diss,we00,nawe01}) are omitted.
\subsection*{Acknowledgments}
We wish to thank 
Alejandro Adem, Jack Mo\-ra\-va, and Yongbin Ruan for
organizing the most memorable ``Workshop 
on Mathematical Aspects of Orbifold String Theory'' in Madison, Wisconsin.
This work is built on the talk presented by the author during that workshop.
It is a summary and extension of the results of 
\cite{nawe00,diss,we00,nawe01}, and the author is grateful to Werner Nahm
for collaboration on part of these works. 
It is  a pleasure to thank
Paul Aspinwall,  David R. Morrison, 
M. Ronen Plesser,  Miles Reid, Falk Rohsiepe, Yongbin Ruan, 
Eric Sharpe, Ber\-nar\-do Uribe, and in particular
Werner Nahm and Andreas Recknagel for helpful discussions and comments,
as well as the referee for very useful criticism.
The author also thanks the  Erwin Schr\"odinger Institute in Vienna
for its  hospitality during the ``Workshop on Mathematical 
Aspects of String Theory'', where the present work was completed.
She was supported by U.S. DOE grant DE-FG05-85 ER 40219, TASK A.
\section{The moduli space $\MM$ of $N=(4,4)$ SCFTs
with $c=6$}\label{modulispace}
Let us briefly describe the structure of the moduli space $\MM$
\cite{asmo94,as96,rawa98,di99,nawe00}. 
There exists a smooth space $\wt\MM$, whose irreducible
components are the unique smooth simply connected spaces 
which by dividing out appropriate groups give the components of $\MM$ 
\cite{th97} and which are determined
entirely by the representation theory of the relevant $N=(4,4)$
superconformal algebra $\AA$. In the following, we will call
$\wt\MM$ the \textsc{smooth covering space} of $\MM$, and analogously
for its components. $\wt\MM$ is not a covering space of $\MM$ in the strict
mathematical sense, since the groups we divide out may act with finite
isotropy.

First, since 
$\AA\supset su(2)_l\oplus su(2)_r$,  one finds
that each holonomy Lie algebra of $\wt\MM$ must contain
$su(2)\oplus su(2)\oplus o(4+\delta)$ for some $\delta\in\N$. Using Berger's 
classification \cite{be55} one concludes  \cite{se88,ce91}
that each ir\-re\-du\-cible component
$\wt\MM^\delta$ of $\wt\MM$ is a Grassmannian of oriented positive definite
four planes in an $\R^{4,4+\delta}$,
\begin{equation}\label{grass}
\wt\MM^\delta \cong \TT^{4,4+\delta} 
\cong O^+(4,4+\delta)/\left(SO(4)\times O(4+\delta)\right).
\end{equation}
Here, for a vector space $W$ with scalar product $\langle\cdot,\cdot\rangle$,
$O^+(W)$ consists of those elements of $O(W)$ which do not interchange
the two components of the space of oriented
maximal positive definite subspaces in 
$W$, and $O(a,b)=O(\R^{a,b})$ etc.
In general,  $\TT^{a,b}$ denotes the Grassmannian of oriented 
maximal positive
definite subspaces in $\R^{a,b}$.
Note that for $b\geq3$, 
$\TT^{4,b}$ is a quaternionic K\"ahler, symmetric
Riemannian manifold of real dimension $4b$ which is simply connected
\cite{wo65}.

Second, our assumptions on SCFTs parameterized by $\wt\MM$ imply that
we can associate an ``elliptic genus'' to each theory in $\wt\MM$. 
In fact, the elliptic genus is an
invariant for each component $\wt\MM^\delta$ and
is given by a theta function of degree $2$ with fixed characteristic
and normalization. There exist only two such functions, namely the
(vanishing) geometric elliptic genus of a complex two torus ($\delta=0$)
and that of
a $K3$ surface ($\delta=16$). By the results of \cite{eoty89}, 
there exist theories of either elliptic genus.

Therefore, the smooth covering space 
of the moduli space of those SCFTs with $c=6$
which provide representations of $\AA$ is
$(\wt\MM^0)^{\oplus N_0}\oplus (\wt\MM^{16})^{\oplus N_{16}}$ with some
nonzero $N_0,\;N_{16}\in\N$. One can prove $N_0=1$ (see, e.g., 
\cite[Ths.\ 7.1.1,\ 7.1.2]{diss}), 
but $N_{16}=1$ remains a widely used conjecture.
However, as we shall explain below, all theories that arise in string theory
and are expected to exhibit connections to geometry are
parameterized either by $\wt\MM^0$ or by a single component of 
$(\wt\MM^{16})^{\oplus N_{16}}$. The corresponding
two-component moduli space $\MM$, for simplicity,
is dubbed \textsc{moduli space of
$N=(4,4)$ SCFTs with central charge $c=6$} nevertheless:
\begin{prop}\label{local}\cite{na86,se88,ce91}
The moduli space $\MM$ decomposes into two  components
$\MM^{tori}=\MM^0$, $\MM^{K3}=\MM^{16}$ 
with smooth covering spa\-ces
$\wt\MM^\delta,\; \delta\in\{0,16\}$ as in \req{grass}. 
The assignment to either component is obtained by the elliptic genus
which associates each theory to the torus or to $K3$.
\end{prop}
As mentioned in the Introduction, the theories in $\MM$ are
important in the context of string theory. In fact \cite{nasu95},
in the case at hand we expect each theory parameterized by 
$\wt\MM=\wt\MM^0\oplus\wt\MM^{16}$ 
to have a nonlinear sigma model realization
describing propagation of strings on some manifold which has the topological
type of a Calabi-Yau
manifold $X$ of complex dimension $2$, i.e.\ a complex two torus
or a K3 surface. 
This justifies our (possible) restriction to the two-component
$\MM$ as in Prop.\ \ref{local}.
Although by definition complex structures exist on $X$, in this
paper we will only occasionally choose a complex structure as an 
auxiliary construction. We rather think of $X$ as a Riemannian 
manifold. This is sufficient for our purposes, since
the parameters of  a nonlinear sigma model 
which describes propagation of strings on $X$ are 
an Einstein metric $g$ on $X$ (possibly in the orbifold limit), and a
so-called B-field $B\in H^2(X,\R)$. The metric $g$ is uniquely
determined by the volume $V\in\R^+$ of $X$ together with the three plane
$\Sigma\subset H^2(X,\R)$ that is invariant under the 
Hodge star operator for $g$, which acts as an involution on $ H^2(X,\R)$
\cite[12.112]{be87}. 
On cohomology, we use the scalar product $\langle\cdot,\cdot\rangle$
which is induced
by the intersection pairing under Poincar\'e duality. Then we have
$H^2(X,\R)\cong\R^{3,3+\delta}$ with $\delta\in\{0,16\}$
as in Prop.\ \ref{local}, and $\Sigma$ is positive definite. In other
words, the parameter space of nonlinear sigma models on $X$ is
\begin{eqnarray}\label{geo}
&&\TT^{3,3+\delta}
\times\R^+\times H^2(X,\R)\nonumber\\
&&\quad\quad\cong O^+(3,3+\delta)/\left(SO(3)\times O(3+\delta)\right)
\times\R^+\times \R^{3,3+\delta}.
\end{eqnarray}
Since \req{geo} parameterizes theories
with elliptic genus given by 
the geometric elliptic genus of $X$ \cite{eoty89}, and
by Prop.\ \ref{local}, the spaces \req{geo} must be isomorphic to the
$\wt\MM^\delta\cong\TT^{4,4+\delta}$. Indeed, for $\delta=16$ this was shown in
\cite{asmo94}, and the same technique works for $\delta=0$. 
The explicit isomorphism depends on the choice of a null
vector $\ups\in\R^{4,4+\delta}$.
For a four plane $x\in\wt\MM^\delta$, since $x$ is maximally positive
definite, $\wh\Sigma:=x\cap\ups^\perp$ is three dimensional. Moreover, we
can find $\xi_4\in x\cap\wh\Sigma^\perp$ with
$\langle\xi_4,\ups\rangle=1$. 
Note that 
$\ups^\perp/\ups\cong\R^{3,3+\delta}$; specify a projection
$pr: \ups^\perp\rightarrow\R^{3,3+\delta}\subset\R^{4,4+\delta}$ 
by choosing another null vector
$\nups\in\R^{4,4+\delta}$ with $\langle\ups,\nups\rangle=1$ such that
the image of $pr$ can  be identified
with $\ups^\perp\cap(\nups)^\perp\subset\R^{4,20}$. Then for
$V:={1\over2}\langle\xi_4,\xi_4\rangle,\; B:=pr(\xi_4-\nups)$,
we find $\xi_4=\nups+B+(V-{\ts{1\over2}}\langle B,B\rangle)\ups$, 
and $\Sigma:=pr(\wh\Sigma)$ obeys $\wh\Sigma=\xi(\Sigma)$ with 
$\xi(\sigma):= \sigma-\langle B,\sigma\rangle\ups$ for $\sigma\in\Sigma$.
In particular, $(\Sigma,\,V,\,B)$ belongs to the product space  
\req{geo} and gives the sigma model data associated to $x$.

The isomorphism we have described depends  on the projection from
$\R^{4,4+\delta}$ $\cong H^{even}(X,\R)$ (without grading) onto 
$\R^{3,3+\delta}\cong H^{2}(X,\R)$ (which fixes the grading). Hence 
$x\in\wt\MM^\delta$ is naturally
interpreted as four plane in $H^{even}(X,\R)$, and $\ups,\,\nups$ as generators
of $H^4(X,\Z),\, H^0(X,\Z)$:
\begin{eqnarray}\label{xi4}
\TT^{4,4+\delta}\cong\wt\MM^\delta\ni \; x & \longmapsto &
(\Sigma,V,B) \;
\in \; \TT^{3,3+\delta}\times\R^+\times H^2(X,\R),\nonumber\\
&x &= \span_\R\left\{ \xi(\Sigma)\,, 
\;\xi_4(V,B)=\nups+B+(V-{\ts{1\over2}}\langle B,B\rangle)\ups \right\},\\
&&\mbox{where for } \sigma\in\Sigma, \;\;
\xi(\sigma):= \sigma-\langle B,\sigma\rangle\ups.\nonumber
\end{eqnarray}
By Poincar\'e duality, the lattices
$H^{even}(X,\Z),\, H^2(X,\Z)$ are even unimodular  of signature
$(4,4+\delta),\, (3,3+\delta)$,  and therefore are isomorphic
to the unique standard even unimodular lattices 
$\Gamma^{4,4+\delta}$ and $\Gamma^{3,3+\delta}$ of the respective signatures
\cite{mi58}. We assume that isomorphisms 
$H^{even}(X,\Z)\cong\Gamma^{4,4+\delta},\, H^2(X,\Z)\cong\Gamma^{3,3+\delta}$
have been fixed and will use $H^{even}(X,\Z),\, H^2(X,\Z)$ synonymously
with $\Gamma^{4,4+\delta},\,\Gamma^{3,3+\delta}$ in the following. In this
sense, $H^{even}(X,\Z)\hookrightarrow H^{even}(X,\R)$ allows
us to use $H^{even}(X,\Z)$ as a reference lattice in 
$H^{even}(X,\R)\cong\R^{4,20}$, where every element of $H^{even}(X,\R)$
is specified by its relative position with respect to $H^{even}(X,\Z)$, 
and analogously for 
$\Gamma^{3,3+\delta}\cong H^2(X,\Z)\hookrightarrow H^2(X,\R)$. 
Mathematically, the latter amounts to a 
choice of marking for $X$, and $\Gamma^{4,4+\delta}\cong 
H^{even}(X,\Z)\hookrightarrow H^{even}(X,\R)$ has been dubbed
\textsc{Mukai marking} \cite{hloy02}. 
Summarizing, we have
\begin{theo}\label{locmod}
The smooth simply connected component $\wt\MM^\delta,\; \delta\in\{0,16\}$, 
of the smooth covering space $\wt\MM$ of the
moduli space $\MM$, which is
associated to a complex two torus or a K3 surface $X$, respectively,
is given by
the Grassmannian of oriented
positive definite four planes in $H^{even}(X,\R)$.

The position of $x\in\wt\MM^\delta$ is specified by its relative position
with respect to the reference
lattice $H^{even}(X,\Z)$. $\wt\MM^\delta$ is
isomorphic to the parameter space \req{geo} of nonlinear sigma models
on $X$. The explicit isomorphism depends on the choice of two 
null vectors $\ups,\,\nups\in H^{even}(X,\Z)$ with 
$\langle\ups,\nups\rangle=1$ as given in \req{xi4}.
For such $x\mapsto(\Sigma,V,B)$
the three plane $\Sigma\subset H^2(X,\R)$, which together with the volume
$V\in\R^+$
determines an Einstein metric
on $X$, is specified by its relative position with
respect to the reference lattice $H^2(X,\Z)$.
\end{theo}
The components $\MM^\delta,\;\delta\in\{0,16\}$, of the moduli space $\MM$
are now obtained from $\wt\MM^\delta,\;\delta\in\{0,16\}$, by dividing out
appropriate discrete groups. These are the groups of 
equivalences of SCFTs with different nonlinear sigma model 
parameters, or in physicists' terminology the \textsc{T-duality groups}.
In fact, by \cite{na86,asmo94,nawe00} 
the appropriate groups are just the orientation preserving lattice
automorphisms of our reference lattices $H^{even}(X,\Z)$. They act
transitively on pairs $(\ups,\nups)$ of null vectors in  $H^{even}(X,\Z)$ with
$\langle\ups,\nups\rangle=1$:
\begin{theo}\label{globmod}
The component $\MM^\delta,\; \delta\in\{0,16\}$,
of the moduli space $\MM$, which is
associated to a complex two torus or a K3 surface $X$, is given by
$$
\MM^\delta
= O^+(H^{even}(X,\Z))\backslash O^+(H^{even}(X,\R))/ SO(4)\times O(4+\delta).
$$
\end{theo}
Summarizing, we have an explicit  description of our moduli
space $\MM$ of $N=(4,4)$ SCFTs with $c=6$: 
The two smooth simply
connected components of its  smooth covering space $\wt\MM$
can be understood
as extensions of the ``geometric'' Teich\-m\"uller spaces of
Einstein metrics (including orbifold limits) on a torus or $K3$ surface
$X$  by the 
additional parameters of 
B-fields $B\in H^2(X,\R)$. In particular, $\MM$ is the moduli space of
such SCFTs with central charge $c=6$ which are representations of $\AA$
and admit nonlinear sigma model descriptions. Let us remark that 
to date, no SCFT with $c=6$
and superconformal algebra $\AA$
has been found not to belong to $\MM$.
The parameters of a SCFT in $\wt\MM$ are
encoded in a positive definite four plane $x\subset H^{even}(X,\R)$
which is specified by its relative position with respect to the lattice
$H^{even}(X,\Z)\subset H^{even}(X,\R)$. Each choice of  
null vectors $\ups,\,\nups\in H^{even}(X,\Z)$ 
with $\langle\ups,\nups\rangle=1$ 
is interpreted as
choice of generators of $H^4(X,\Z),\,H^0(X,\Z)$, respectively. It 
fixes the projection $H^{even}(X,\R)\rightarrow H^{2}(X,\R)$ and thereby
a \textsc{geometric interpretation} $(\Sigma,V,B)$ of $x$.
If $B=0$ in such a geometric interpretation, then $x$ is the $+1$
eigenspace  in $H^{even}(X,\R)$  of the Hodge star operator which 
corresponds to the Einstein metric given by $(\Sigma,V)$.
All equivalences of SCFTs in $\wt\MM$ are lattice automorphisms
in $O^+(H^{even}(X,\Z))$.

In the Introduction, we have mentioned that $\MM$ is a partial completion
of the actual moduli space of SCFTs we are interested in. Namely,  
$\MM^{K3}$ contains  points which
do not correspond to well-defined SCFTs \cite{wi95}.
They form subvarieties of $\MM^{K3}$ with  at least
complex codimension $1$ \cite{agm94a}. 
These ill-behaved theories,
however, will not be of relevance for the discussion below.
\section{Orbifold conformal field theories on $K3$}\label{orbifolds}
In the previous section, we have described the moduli
space $\MM$ of $N=(4,4)$ SCFTs with $c=6$ as algebraic space.
Every theory in its toroidal component $\MM^{tori}$ can be explicitly 
constructed as a
toroidal SCFT \cite{na86}. Anything but an analogous statement is true for  
$\MM^{K3}$, however,  due to the fact that no smooth
Einstein metric on $K3$ is known explicitly. We can only construct a finite
subset of theories in $\MM^{K3}$, 
which is given by Gepner models and orbifolds 
thereof (\textsc{Gepner type models}), and CFTs in 
lower dimensional subvarieties
obtained by an orbifold procedure from  theories in appropriate
subvarieties of $\MM^{tori}$.
The description of these subvarieties of $\MM^{K3}$ is the object
of the present section.

Given a SCFT $\CC$ and a finite group $G$ that acts on its Hilbert space, 
under certain additional
assumptions on this action one can construct 
a well-defined \textsc{orbifold conformal field theory} $\CC/G$. 
It is obtained by projecting
onto $G$ invariant representations of the superconformal algebra  and
adding so-called \textsc{twisted representations}\footnote{See
Sect.\ \ref{cftvsgeo} for further details.}, 
each of which is
nonlocal with respect to some representation in the original theory $\CC$.

In the case of interest to us, where $\CC=\CC_T$ is a fixed toroidal
SCFT corresponding to some $x_T\in\MM^{tori}$, all assumptions on 
the $G$ action are fulfilled
if we can find a geometric interpretation $(\Sigma_T,V_T,B_T)$
for $x_T$ on a $G$ symmetric torus $T$ with metric given by $(\Sigma_T,V_T)$
such that $B_T\in H^2(T,\R)^G$ and the action of $G$ is induced by
the geometric action on $T$. 
These data are assumed to be fixed in the 
following. We suppose that 
$G$ acts non-trivially on $T$ and
does not contain non-trivial translations. Moreover, we assume that 
all singularities $s\in\SS\subset T/G$ can be minimally
resolved to obtain a $K3$ surface 
$p:X=\wt{T/G}\rightarrow T/G$, which in particular implies
$G\subset SU(2)$.  Such $G$ actions have been classified 
\cite{fu88}, and the relevant groups
are cyclic, binary 
dihedral or tetrahedral, respectively\footnote{By \cite{fu88},
two different actions of $\wh D_4$ with different fixed point structures
exist. Note that for our groups we are using notations that
are common in singularity theory, where $\C^2/\wh D_n$ has a $D_n$
type singularity at the origin and $\wh D_n$ has order $4(n-2)$.
The binary tetrahedral group $\wh\T$ has order $24$ and produces an
$E_6$ type singularity at the origin of $\C^2/\wh\T$.}:
\begin{equation}\label{notations}
\Z_N,\; N\in\{2,3,4,6\}, \quad
\wh D_n,\; n\in\{4,5\}, \quad \wh\T.
\end{equation}
As to notations, let 
$\pi:T\rightarrow X$ denote
the induced rational map of degree $|G|$ which
is well defined away from the fixed points of $G$. Each singularity
$s\in\SS\subset T/G$ 
is of ADE type, such that the intersection matrix for the
irreducible components
of the exceptional divisor $p^{-1}(s)$  is given by the negative
of the Cartan matrix of the ADE group corresponding to the singularity
$s$. The Poincar\'e duals of the $(n_s-1)$
components of $p^{-1}(s)$ are denoted\footnote{Many of the objects
introduced below have hats for historical reasons, since in 
\cite{nawe00} we mainly used the unhatted objects.
In order to keep our notations consistent, we keep these hats, and we
apologize to the reader for the ugly notation.}
$\wh E_s^{(l)},\; l\in\{1,\dots,n_s-1\}$. Moreover, the $\Z$-span of these cocycles
is denoted $\wh{\EE}_s\subset H^2(X,\Z)$, and 
$\wh\EE:=\oplus_s\wh\EE_s$. In writing 
$X=\wt{T/G}$ we mean the orbifold limit of $K3$, i.e.\ $X$ is equipped with the
metric induced by the flat metric $(\Sigma_T,V_T)$
on $T$ which assigns volume zero to
each cycle corresponding to an $\wh E_s^{(l)}\in\wh\EE$.

By calculating the elliptic genus it is not hard to
check that the orbifold CFT $\CC_T/G$ belongs
to the $K3$ component $\MM^{K3}$ of the moduli space. In the following,
we will in fact assume that it has a geometric interpretation 
on 
$X=\wt{T/G}$. How such a geometric interpretation $(\Sigma,V,B)$ 
is consistently obtained shall be
explained in two steps. First, the metric on $X$ has to be
specified by giving
the volume $V$ of $X$ and $\Sigma\subset H^2(X,\R)$. Second, we 
need to determine the B-field $B\in H^2(X,\R)$ of $\CC_T/G$.

For the first step, the ``geometric'' part of the problem, we have $V_T=|G|V$,
and we know that $\Sigma$ will be 
specified by its relative position with respect
to $H^2(X,\Z)$. Since the relative position of $\Sigma_T$ with respect
to $H^2(T,\Z)^G$ is known, the strategy is to determine the embedding
$\pi_\ast H^2(T,\Z)^G\hookrightarrow H^2(X,\Z)$.  This can be done by 
generalizing methods due to Nikulin, who in \cite{ni75}
considered the case $G=\Z_2$.
To this end, note that $\pi_\ast H^2(T,\Z)^G\perp\wh\EE$
is a sublattice of maximal rank in $H^2(X,\Z)$. Moreover, by \cite{in76}
one has
$\pi_\ast H^2(T,\Z)^G\cong H^2(T,\Z)^G(|G|)$, where given
a lattice $\Gamma$, by
$\Gamma(n)$ one denotes the lattice which agrees with $\Gamma$ as a 
$\Z$ module but has quadratic form scaled by a factor of $n$.\footnote{In 
\cite{nawe00,diss,we00,nawe01} we have
implement this by identifying $\pi_\ast H^2(T,\Z)^G 
\cong
\{ \sqrt{|G|}\kappa\mid\kappa\in H^2(T,\Z)^G\}$, where for 
$\kappa\in H^2(T,\Z)^G$ we kept on using the original scalar product. 
That notation is convenient for calculations but introduces many square
roots in the formulae, which we will avoid in this note.}
The key observation is
that it suffices to find the maximal primitive sublattices
$K_{|G|}\supset \pi_\ast H^2(T,\Z)^G,\; \wh\Pi_{|G|}\supset\wh\EE$
in $H^2(X,\Z)$ and apply the following Th.\ \ref{lattemb}. It 
allows to describe the lattice $ H^2(X,\Z)$ in terms
of the sublattices $K_{|G|},\;\wh\Pi_{|G|}$:
\begin{theo}\cite[Prop.1.6.1]{ni80b}, \cite[\para1]{ni80}\label{lattemb}
Let $\Lambda$ denote  a primitive nondegenerate sublattice of an even
unimodular lattice $\Gamma$ and $\Lambda^*$ its dual, with
$\Lambda\hookrightarrow\Lambda^*$ by  use of the quadratic form on $\Lambda$. 
Then the embedding $\Lambda\hookrightarrow\Gamma$
with $\Lambda^\perp\cap\Gamma\cong \VV$ is specified by an isomorphism 
$\gamma:\Lambda^\ast/\Lambda\rightarrow \VV^\ast/\VV$, such
that the induced quadratic forms  
$q_\Lambda:\Lambda^*/\Lambda\rightarrow\Q/2\Z,\;
q_\VV:\VV^*/\VV\rightarrow\Q/2\Z$ obey
$q_\Lambda=-q_\VV\circ\gamma$. Moreover,
$$
\Gamma\cong\left\{ (\lambda,\vee)\in \Lambda^\ast\oplus \VV^\ast\mid
\gamma(\qu\lambda)=\qu{\vee} \right\},
$$
where $\qu l$ denotes the projection of $l\in L^\ast$ onto $L^\ast/L$.
\end{theo}
In \cite{ni75} Nikulin showed that
for $G=\Z_2$ one has  $K_2\cong 
H^2(T,\Z)(2)$, and  $\wh\Pi_2$
is the \textsc{Kummer lattice} \cite{pss71}. 
Let $\mu_1,\dots,\mu_4$ denote generators of $H^1(T,\Z)$ and
$Q_{i,j}^k,\; i,j,k\in\{1,\dots,4\}$, the $\Z_2$ invariant representatives of
the Poincar\'e dual of $2\mu_i\wedge\mu_j$ with 
$\SS\cap (Q_{i,j}^k/\Z_2)\neq\empty$. 
Then one finds
\begin{eqnarray}\label{split}
H^2(X,\Z) &=& \span_\Z\left\{ 
\pi_\ast(\mu_i\wedge\mu_j)\in K_2;\quad\epsilon\in\wh\EE; \right.\nonumber\\
&&\left.
\quad\quad
{\ts{1\over2}}\pi_\ast(\mu_i\wedge\mu_j)
\;+\;\smash{{\ts{1\over2}}\hspace*{-1.5em}
\sum_{s\in\SS\cap (Q_{i,j}^k/\Z_2)}}\hspace*{-1.5em}
\wh E_s,\quad i,j,k\in\{1,\dots,4\}\right\}.
\vphantom{\sum_{s\in\SS\cap Q_{i,j}}}
\end{eqnarray}
In \cite{we00}, a description analogous to \req{split} was given
for all cases under discussion. In particular, it suffices to determine
the pairs $\kappa+\epsilon\in H^2(X,\Z)$ with 
$\kappa\in K_{|G|}^*$ and $\epsilon\in\wh\Pi_{|G|}^*$
but nonzero $\qu\kappa,\;\qu\epsilon$. Their 
$\wh\Pi_{|G|}^*$ 
contributions $\qu\epsilon$ are always given by linear combinations of
${1\over|G|}\wh E_s^{(l)}$ with $s\in\SS\cap (Q/G)$ for an appropriate 
representative $Q$ of
the Poincar\'e dual of $\pi^*(\kappa)$. From 
\cite[Prop.2.1]{we00} one obtains\footnote{For cyclic groups
$G=\Z_N,\; N\in\{3,4,6\}$, $\wh\Pi_N$ was first found in \cite{be88}.}
all $K_{|G|},\;\wh\Pi_{|G|}$ as well as
the embeddings $\pi_\ast H^2(T,\Z)^G\hookrightarrow H^2(X,\Z)$.
Hence, we can specify the relative 
position of $\Sigma:=\pi_\ast\Sigma_T$ with respect to $H^2(X,\Z)$, 
and $(\Sigma,V)$ specifies the Einstein metric on $X$.

The second step in the determination of $(\Sigma, V, B)$ uses the same idea
as the first one. In \cite{we00} we show 
that the images $\pi_\ast(\ups),\,\pi_\ast(\nups)$ of $\ups,\,\nups$
in $H^{even}(X,\Z)$ generate a primitive sublattice. Moreover, 
the maximal primitive sublattice $\wh K_{|G|}$ of 
$\pi_\ast H^{even}(T,\Z)^G$ in $H^{even}(X,\Z)$ obeys
$$
\wh K_{|G|}^\ast / \wh K_{|G|}
\cong K_{|G|}^\ast / K_{|G|} \times \Z_{|G|}^2
\cong \wh\Pi_{|G|}^*/\wh\Pi_{|G|} \times \Z_{|G|}^2.
$$
Hence Th.\ \ref{lattemb} implies that
$\wh K_{|G|},\, \wh\Pi_{|G|}$ cannot be embedded
in $H^{even}(X,\Z)$ as orthogonal sublattices. 
Rather, instead of $\pi_\ast(\ups)$ and $\pi_\ast(\nups)$ with
$\langle\pi_\ast(\ups),\pi_\ast(\nups)\rangle=|G|$, appropriate
generators of $H^4(X,\Z)$ and $H^0(X,\Z)$ are
\begin{equation}\label{newmlattice}
\wh\upsilon:=\pi_\ast(\upsilon), \quad\quad
\wh\upsilon^0 := {\ts{1\over|G|}}\pi_\ast(\upsilon^0)
- {\ts{1\over|G|}}\wh B_{|G|}-\wh\upsilon, 
\quad \wh B_{|G|}\in\wh\Pi_{|G|},
\end{equation}
and $\wh B_{|G|}$ is uniquely determined up to  
irrelevant lattice automorphisms \cite[Lem.\ 3.2]{we00}. This  gives 
$\pi_\ast H^{even}(T,\Z)^G\hookrightarrow H^{even}(X,\Z)$; since the
relative position of $x_T$ with respect to $H^{even}(T,\Z)^G$ is known,
it allows to read off the desired geometric interpretation of the
orbifold CFT corresponding to $x=\pi_\ast x_T\in\MM^{K3}$. In particular,
$B$ is obtained\footnote{For $G=\Z_2$, the correct B-field 
was first found in \cite{as95}; using D-geometry, it was determined
for all cyclic groups in \cite{do97,blin97}; together with W.~Nahm
in \cite{nawe00} we rederive
it for $\Z_2$ and $\Z_4$.} 
by rewriting the vector $\xi_4(V_T,B_T)$ of \req{xi4}
in terms of  $\wh\ups,\,\wups$ instead of $\ups,\,\nups$:
\begin{prop}\label{z4emb}\cite[Th. 3.3]{we00}
Let $(\Sigma_T,V_T,B_T)$ denote a geometric interpretation of a toroidal 
SCFT $x_T\in\MM^{tori}$ on the torus $T$ that admits a $G$ symmetry,
$G\subset SU(2)$ not containing non-trivial
translations; all possible $G$  are listed
in \req{notations}. Then its image
$x\in\MM^{K3}$ under the
$G$ orbifold procedure has geometric interpretation
$(\Sigma,V,B)$ where $\Sigma=\pi_\ast\Sigma_T$,  $V={V_T\over {|G|}}$,
and $B={1\over|G|}\pi_\ast(B_T)+{1\over|G|}\wh B_{|G|}$. Here,
$\wh B_{|G|}\in\wh\Pi_{|G|}$, and
for each $G_s\subset G$ type
singularity $s\in\SS$ and $l\in\{1,\dots,n_s-1\}$, 
$\langle\wh B_{|G|},\wh E_s^{(l)}\rangle$ is 
the $|G\colon G_s|$--fold coefficient of 
$\wh E_s^{(l)}$ in the highest root
of $\wh\EE_s$.
\end{prop}
Summarizing, for the $G$ orbifold of a toroidal
theory $\CC_T$ in $\MM^{tori}$ the choice of geometric interpretation on
the orbifold limit $\wt{T/G}$ of $K3$ induces a certain nonzero
fixed B-field ${1\over|G|}\wh B_{|G|}$
in direction of the exceptional divisor of the blow
up, as predicted in \cite{as95}. This B-field can be determined explicitly
by classical geometric considerations. In particular, \req{newmlattice}
implies
\begin{equation}\label{mumu}
|G|\upsilon^0 =
\pi^\ast\left(|G|\,\wh\upsilon^0 + \wh B_{|G|} + |G|\wh\ups\right),
\end{equation}
which allows for an interpretation \cite{we00}
in terms of the classical McKay
correspondence \cite{mk80,mk81}. 
Namely, we interpret \req{mumu} as
an equation of Mukai vectors for vector bundles on $T,\,X$. More
precisely, $|G|\nups$ corresponds to a flat bundle of rank $|G|$ on $T$
that naturally carries the regular representation of $G$ on the
fibers, yielding a $G$ equivariant flat bundle. By the results 
in \cite{bkr99}, there is a corresponding bundle on $X$ whose 
Mukai vector should be given by the argument of $\pi^*$ in \req{mumu}.
In particular, $\wh B_{|G|}$ must be the first Chern class of that bundle,
which receives a contribution $|G\colon G_s|\,\wh B_{|G|}^s$
from each $G_s\subset G$ type singularity
$s\in\SS$ that is determined by
the classical McKay correspondence \cite{gsve83,kn85,arve85}:
According to the decomposition $\rho_s=\sum_l m_s^{(l)}\rho^{(l)}$
of the regular representation $\rho_s$ of $G_s$ into irreducible
ones, we have $\wh B_{|G|}^s=\sum_l m_s^{(l)} (\wh E_s^{(l)})^\ast$, where 
$\sum_l m_s^{(l)} \wh E_s^{(l)}$ is the highest root of $\wh\EE_s$, and 
$\{ (\wh E_s^{(l)})^\ast\}\subset\wh\EE_s\otimes\Q$ 
denotes the dual basis of the fundamental system
$\{ \wh E_s^{(l)}\}$ of $\wh\EE_s$. 
This is in exact agreement with Prop.\ \ref{z4emb}. For cyclic $G$ we can 
calculate the entire Mukai vector that is expected on the right hand side
of \req{mumu} and find agreement.
\section{Recreational interlude on mirror symmetry}\label{mirror}
Since 
in Sect.\ \ref{example} we will compare various approaches to mirror symmetry
on $K3$ in a particular example, 
the present section is devoted to a sketchy overview of basic ideas of mirror 
symmetry. We apologize for the inevitable incompleteness but refer the
reader to the literature for details (see e.g.\ \cite{coka99}). 

We view mirror symmetry as an equivalence of $N=(2,2)$ SCFTs
induced by the outer automorphism of the left handed $N=2$ superconformal
algebra which inverts the sign of the $U(1)$ current and interchanges
the two supercharges \cite{di87,lvw89}. If the equivalent SCFTs admit 
(at least  approximate) geometric interpretations as nonlinear sigma models
on different Calabi-Yau manifolds $X,\,X^\prime$, 
mirror symmetry must induce a ``nonclassical
duality'' between geometrically unrelated manifolds, an observation that
has had striking impact on both mathematics and physics. 
Meanwhile, it has become a well-known slogan that mirror symmetry interchanges
complex and (quantum corrected) complexified K\"ahler moduli of $X,\,X^\prime$.

The first explicit construction of mirror dual SCFTs was given by 
Greene and Plesser in 
\cite{grpl90}. For each Gepner type  model $\CC/H$ obtained as Abelian
$H$ orbifold
from a Gepner  model $\CC$ with  
central charge $c=3d,\; d\in\N\,$, they determine
an Abelian group $H^*$ of symmetries\footnote{We call $H^*$ the 
\textsc{Greene/Plesser (GP) group}. 
For further details, see the proof of 
Prop.\ \ref{msself}.}  of $\CC$
such that the orbifold $\CC/H^*$ is the mirror of $\CC/H$.
A specific subgroup $G$ of
the symmetry group of a Gepner type model can be used 
to characterize the family of Calabi-Yau manifolds
which admit an algebraic $G$ action \cite{ge87,ge88}. These 
Calabi-Yau manifolds should be lowest order approximations to string 
vacua constructed from the corresponding Gepner type model. This enabled
Greene and Plesser to give a geometric meaning to their
version of mirror symmetry as a duality between families of Calabi-Yau
manifolds, with highly non-trivial verification found in \cite{cogp91,mo93}.

The idea to characterize families of Calabi-Yau manifolds by their
algebraic automorphisms translates nicely into the construction of 
families of Calabi-Yau hypersurfaces in toric varieties. Indeed, the ``toric
description'' of mirror symmetry states that the family of Calabi-Yau toric
hypersurfaces corresponding to the reflexive polyhedron $\Delta$ has a
mirror dual corresponding to the dual polyhedron $\Delta^*$
\cite{ba94}.

In Sect.\ \ref{modulispace} we have argued that each theory in our moduli
space $\MM$ admits nonlinear sigma model realizations on a complex two torus
or a $K3$ surface $X$. 
Moreover, from 
Th.\ \ref{globmod} we know that all equivalences of theories parameterized
by $\wt\MM$ are given by lattice automorphisms in $O^+(H^{even}(X,\Z))$.
It is therefore natural to ask which of these automorphisms should describe
mirror symmetry. The question is delicate since all theories in $\wt\MM$
have enhanced supersymmetry beyond $N=(2,2)$, so to address the outer
automorphism on the superconformal algebra $\AA$ which inverts the sign
of the left handed $U(1)$ current requires the choice of a Cartan torus in 
$su(2)_l\subset\AA$. In terms of geometric interpretations this corresponds
to the choice of a complex structure within the $\S^2$ of complex structures
compatible with a given Einstein metric.
As mentioned after Prop.\ \ref{local}, such a choice has not been made in 
our constructions up to now.

A solution to this problem that is closely related 
to the results on mirror symmetry
in the context of toric geometry has been proposed by Aspinwall and Morrison
\cite{asmo94}. To this end, let us concentrate on $\MM^{K3}$. We consider
a family $\{x_t\}\subset\wt\MM^{K3}$ of so-called \textsc{$M$ polarized}
theories, where $M\subset H^2(X,\Z)$ is a primitive sublattice of signature
$(1,\rho-1),\; \rho\geq1$. Namely, it is assumed that for each $x_t$
a geometric interpretation $(\Sigma_t,V_t,B_t)$, and also a compatible
complex structure
have been chosen such that $M$ can be embedded as primitive sublattice 
into the 
corresponding Picard lattice,
and then $B_t\in M\otimes\R\,$. The family
of $\check M$ polarized theories is a
mirror dual family iff there is an embedding of $M\perp\check M$ as
sublattice of maximal rank into the unique  standard
even unimodular lattice 
$\Gamma^{2,18}\subset H^2(X,\Z)$ of 
signature $(2,18)$. 
This construction has been discussed in detail by
Dolgachev \cite{do96}, where he also  explains 
that  Arnol'd's strange duality \cite{ar74,doni77,pi77} actually is the
oldest version of mirror symmetry ever investigated (see also 
\cite[\para5.8]{gi92}).
The existence and uniqueness
of $\check M$, however, cannot be proven in general,
since an embedding $M\hookrightarrow\Gamma^{2,18}\subset H^2(X,\Z)$
might not exist (uniquely).

The characterization of 
a family of Calabi-Yau manifolds by its generic Picard lattice
is related to a characterization by its generic
algebraic automorphisms. It is therefore natural to expect that this approach
to mirror symmetry should have a toric cousin. 
Indeed, for any family of hypersurfaces in a toric variety corresponding to 
a reflexive polyhedron $\Delta$ we can determine the generic Picard lattice
$Pic(\Delta)$. Moreover,  
if in the $K3$ case all divisors
in $Pic(\Delta)$ are toric,
$Pic(\Delta^*)\cong\check{Pic}(\Delta)$. 
This, however, is not the case in general. 
Instead, let $Pic_{tor}(\Delta)\subset Pic(\Delta)$ denote the sublattice
of toric divisors in $Pic(\Delta)$, then Rohsiepe found \cite{ro01a,ro01b}
$Pic(\Delta^*)\cong\check{Pic}_{tor}(\Delta)$ and 
$Pic_{tor}(\Delta^*)\cong\check{Pic}(\Delta)$. Therefore, one arrives at the
following more refined statement about mirror symmetry on $K3$:
\begin{prop}\cite[Prop.4.1]{ro01a}\label{falksms}
Let $\{x_t\}\subset\wt\MM^{K3}$ denote the family of theories with
geometric interpretation $(\Sigma_t,V_t,B_t)$, such that $(\Sigma_t,$ $V_t)$
specifies the family of hypersurfaces in a toric variety corresponding
to the reflexive polyhedron $\Delta$, 
and $B_t\in Pic_{tor}(\Delta)\otimes\R\,$.
Then there is a mirror family with geometric interpretation 
$(\check\Sigma_t,\check V_t,\check B_t)$ with $(\check\Sigma_t,\check V_t)$
corresponding to the dual polyhedron $\Delta^*$ and 
$\check B_t\in Pic_{tor}(\Delta^*)\otimes\R=\check{Pic}(\Delta)\otimes\R\,$.
\end{prop}
All geometric
constructions of mirror symmetry 
mentioned so far refer to appropriate
families of Calabi-Yau manifolds. In contrast, on the level of SCFTs
the Greene/Plesser construction is a point to point map on $\wt\MM$. We
will argue that in favorable situations it is also possible to give a
geometric point to point map for mirror symmetry, as expected from
Th.\ \ref{globmod}. In fact, Sect.\ \ref{example} is devoted to the
discussion of an example where all versions of
mirror symmetry on $K3$ can be calculated and compared, and we will show
that they are equivalent there.

The basic idea goes back to Vafa and Witten \cite{vawi95}; 
consider a toroidal SCFT $\CC_T$ with sigma model realization on a 
$d$ dimensional complex torus $T^{2d}$. Projection onto the real part
of each complex coordinate gives a torus fibration
$T^{2d}\rightarrow T^d$ over a $d$ dimensional real torus. In \cite{vawi95}
it is shown that fiberwise T-duality in this fibration induces the 
mirror automorphism on the superconformal algebra of $\CC_T$. Moreover, for
a geometric symmetry $G$ of $T^{2d}$ which respects the fibration 
and supersymmetry, one also obtains
a mirror map for the orbifold $\CC/G$. This idea has been extended in 
\cite{syz96} to the celebrated Strominger/Yau/Zaslow (SYZ) conjecture
which is supposed to hold for more general fibrations. It
should be noted that for a generic member of an $M$ polarized 
family of models in 
$\wt\MM^{K3}$ the assumptions necessary for the SYZ mirror 
construction are fulfilled iff one can 
construct the mirror family \`a la Dolgachev \cite[Cor.1.4]{grwi97}.

In \cite{nawe01} together with W.~Nahm
we have used Vafa and Witten's approach to give an
explicit construction of the mirror automorphism 
$\gamma_{MS}\in O^+(H^{even}(X,\Z))$ for $\Z_N$ orbifold CFTs in 
$\wt\MM^{K3}$. First, using \cite{na00b} for $\wt\MM^{tori}$
we determine the appropriate lattice automorphism
in $O^+(H^{even}(T,\Z))$ that is induced by fiberwise T-duality of
$T\rightarrow T^2$. By making use of the $\Z_N$ orbifold constructions
discussed in Sect.\ \ref{orbifolds} we extend this
automorphism to an element of $O^+(H^{even}(X,\Z))$ with 
$X=\wt{T/\Z_N},\; N\in\{2,3,4,6\}$. 
In other words, with a suitable complex structure 
\cite{hala82} we  determine the mirror map which is
induced by fiberwise T-duality in an elliptic fibration
$X\rightarrow\P^1$ of the orbifold limit $X$ of $K3$. 
In particular, we calculate the explicit action
on the (non-stable) singular fibers of this fibration.
 
The extension
of the lattice automorphism in $O^+(H^{even}(T,\Z))$ to an element of
$O^+(H^{even}(X,\Z))$ is possible since by the discussion of 
Sect.\ \ref{orbifolds} we know the explicit embeddings
$\pi_\ast H^{even}(T,\Z)^{\Z_N}\hookrightarrow H^{even}(X,\Z)$.
First recall that for $s\in\SS$, $l\in\{1,\dots,n_s-1\}$ the 
$\wh E_s^{(l)}\in H^2(X,\Z)$ are not orthogonal to 
$\wh K_N\supset\pi_\ast H^{even}(T,\Z)^{\Z_N}$. By Prop.\ \ref{z4emb}
\begin{equation}\label{ehat}
E_s^{(l)}
:=\wh E_s^{(l)}- {\ts{1\over{n_s}}}\wh\ups
\end{equation}
are the orthogonal projections onto $\wh K_N^\perp$ in
$H^{even}(X,\R)$. Analogously, $\Pi_N$ 
denotes the orthogonal projection
of $\wh\Pi_N$ onto $\wh K_N^\perp\cap H^{even}(X,\R)$, and 
similarly for $\EE_s,\; \EE$. 
The mirror map must act as lattice automorphism on $\Pi_N$.
In each case we have a description for $H^{even}(X,\Z)$ 
analogous to \req{split}. Knowing the mirror map on $\wh K_N$
and applying it to lattice vectors of type
${1\over N}\pi_\ast(\kappa)+{1\over N}\sum \wh E_s^{(l)},\; 
{1\over N}\pi_\ast(\kappa)\in K_N^*,\;
\wh E_s^{(l)}\in\wh\Pi_N$, already determines the mirror map on $\Pi_N$
up to automorphisms which are entirely under control \cite{nawe01}.
For brevity, here we only state the formula for the action of
mirror symmetry $\gamma_{MS}$
on the Kummer lattice $\Pi_2$:
\begin{prop}\label{geoms}
Consider a $\Z_N$ orbifold CFT on $K3$ constructed from a toroidal
CFT with $\Z_N$ symmetric torus and vanishing B-field. 
For the $\Z_2$ orbifold limit of $K3$, we also assume the underlying torus to
be orthogonal. Here, we
use\footnote{As usual, 
$\F_p$ with $p$ prime denotes the unique
finite field with $p$ elements.}
$\SS\cong\F_2^4$ to label the generators of $\EE$ by $\F_2^4$,
where the first two coordinates correspond to the fiber coordinates
of the torus fibration. 
Then the version of mirror symmetry which is induced by fiberwise T-duality
on the underlying toroidal theory acts on $\Pi_2$ by
$$
\forall\; (I,J,K,M)\in\F_2^4:\quad
\gamma_{MS}(E_{(I,J,K,M)}) 
= {\ts{1\over2}} \sum_{i,j\in\F_2} (-1)^{iI+jJ}E_{(i,j,K,M)}.
$$
Similar formulas for the other $\Z_N$ orbifold limits of $K3$ are given in
\cite[(20)]{nawe01}. $\gamma_{MS}$ is 
uniquely determined up to certain permutations
of $\SS$ and automorphisms
of $\Pi_N$ which are parameterized by $b\in\Pi_N/\EE$ and act on 
the B-field by a shift by $b$.
\end{prop}
\section{Conformal field theory versus Geometry}\label{cftvsgeo}
In the previous sections, we have mostly used geometric arguments to
describe the moduli space $\MM$ of $N=(4,4)$ SCFTs with $c=6$,
its orbifold subvarieties and mirror symmetry for them. In the present
section we will discuss corresponding results in CFT language in
order to clarify the direct link between geometry and SCFT.

In Sect.\ \ref{orbifolds} we have only briefly  mentioned the construction of
an orbifold SCFT $\CC/G$ from a given theory $\CC$
by projecting onto $G$ invariant representations
of the superconformal algebra and adding twisted ones. The
lowest weight 
states of the latter are called \textsc{twisted ground states} and
generate a finite dimensional subspace $\TT$ of the Hilbert space of 
$\CC/G$. A basis of $\TT$ can be labeled $T_{s}^{[g]}$, where
$[g],\;g\in G$, denotes a
non-trivial conjugacy class in $G$,
and $s\in\SS_{[g]}$ accounts
for degeneracies. If $\CC$ has a nonlinear sigma model realization on some
Calabi-Yau manifold $Y$ and the $G$ action is induced by a geometric 
action on it, then  $\SS_{[g]}$ consists of the
$G$ orbits in $Y$ that are pointwise fixed by $g$.

Hence for the orbifold CFTs on $K3$ orbifold limits $X$ 
of Sect.\ \ref{orbifolds} the twisted ground states
can be labeled $T_s^l,\; s\in\SS,\; l\in\{1,\dots,n_s-1\}$. Their $1:1$
correspondence to components
$\wh E_s^{(l)}$ of exceptional divisors in the blow-up will be worked out
in detail below. In fact, there is a standard scalar product 
$\langle\cdot,\cdot\rangle$ on the Hilbert space of each CFT with respect to
which all states in the twisted representations are orthogonal to each state
in the $G$ invariant part of the original theory's Hilbert space. It
is therefore even more natural to expect a $1:1$ correspondence between
twisted ground states and the projections $E_s^{(l)}$ onto
$\wh K_{|G|}^\perp$, which for the $\Z_N$ orbifold CFTs will indeed be
established below. We normalize the twisted ground states such that
with $\zeta_n$ denoting a fixed primitive $n^{th}$ root of unity
\begin{eqnarray}\label{norm}
\forall s,\;s^\prime\in\SS,
&&\!\!\!\!\!\!\!\!\!
l^{(\prime)}\in\{1,\dots,n_{s^{(\prime)}}-1\}: 
\nonumber\\
&&\left\langle T_s^l, T_{s^\prime}^{l^\prime}\right\rangle
= \delta_{s,s^\prime}\delta_{l,l^\prime}
\left( 2-\zeta_{n_s}^l-\zeta_{n_s}^{-l}\right).
\end{eqnarray}
From now on, we will restrict to the case of $\Z_N$ orbifold CFTs 
on $K3$ discussed in Sects.\ \ref{orbifolds}, \ref{mirror}. The underlying
toroidal theory $\CC_T$ is assumed to have a geometric interpretation on 
a $\Z_N$ symmetric
four torus $T=\R^4/\Lambda$ with vanishing B-field. Here, 
$\Lambda\subset\R^4$ is a nondegenerate lattice of rank $4$, and its dual
is denoted $\Lambda^*$. In the $\Z_2$ case we assume $\Lambda$ to be
orthogonal.
The toroidal theory $\CC_T$ then possesses so-called
\textsc{vertex operators} which are parameterized by $\Lambda^*\oplus\Lambda$
and also act on the twisted ground states of $\CC_T/\Z_N$.  As to notations, 
we use $\SS\hookrightarrow {1\over N}\Lambda/\Lambda$ and denote by
$\theta$ the generator of the geometric $\Z_N$ action on $T$
which naturally acts on $\Lambda$ as well. The action of the vertex operators
then induces the following representation
$W$ of $\Lambda^*\oplus\Lambda$ on $\TT$ by restriction to leading order
terms in the OPE
(see \cite[Sect.\ 5]{nawe01} 
for details):
\begin{eqnarray}\label{weyl}
\forall\, (\mu,\lambda)\in\Lambda^*\oplus\Lambda;
&&\!\!\!\!\!\!\!\!\!\forall\, s\in\SS, l\in\{1,\dots,n_s-1\}: \nonumber\\
W(\mu,\lambda) T_s^l &=& \zeta_N^{l\mu(Ns)} T_{s^\prime}^l ,
\quad
s^\prime = s+(\id-\theta)^{-1}\lambda  
= s-{\ts{1\over n_s}} \sum_{k=1}^{n_s-1} k\theta^k\lambda.
\end{eqnarray}
In particular, we have
\begin{eqnarray*}
\forall\, q,q^\prime\in\Lambda^*\oplus\Lambda,&&
\!\!\!\!\!\!\!\!\!\forall\, s\in\SS, l\in\{1,\dots,n_s-1\}: \\
W(q^\prime)W(q) T_s^l 
&=& \zeta_{n_s}^{l\phi_{n_s}(q,q^\prime)} W(q)W(q^\prime) T_{s}^l ,\\
&&\;
\phi_n( (\mu,\lambda),(\mu^\prime,\lambda^\prime) )
:= \sum_{k=1}^n k\left( 
\mu\theta^k\lambda^\prime-\mu^\prime\theta^k\lambda\right).
\end{eqnarray*}
In other words, $W$ is a Weyl algebra representation\footnote{Equation
\req{weyl} also shows that $W$ is a \textsc{projective representation},
a term which is more common in the modern physics literature.} 
of $\Lambda^*\oplus\Lambda$ on $\TT$, where $\Lambda^*$ acts diagonally by 
introducing phases, and $\Lambda$ acts by translation on the base point
$s\in\SS$.

The action of (fiberwise) T--duality on the vertex operators
of the toroidal theory $\CC_T$ is given by an exchange of rank $2$
sublattices of $\Lambda^*$ and $\Lambda$. By using
the above Weyl algebra representation and re-diagonalizing appropriately
it is therefore possible to
derive the effect of (fiberwise) T--duality on twisted ground states in
$\TT$. This was performed in joint work with W.~Nahm in \cite{nawe01}. 
Again, for brevity
we only state the result for $G=\Z_2$ explicitly:
\begin{prop}\label{cftms}
Consider a $\Z_N$ orbifold CFT on $K3$ constructed from a toroidal
SCFT with $\Z_N$ symmetric torus and vanishing B-field. For the $\Z_2$
case we also assume the underlying torus to be orthogonal.
Then the action  of fiberwise
T-duality on the underlying toroidal theory induces a $\Z_N$
type fiberwise Fourier transform $F$ on the twisted ground states in 
$\TT$. Explicitly, for $\Z_2$ we have
$$
\forall\; (I,J,K,M)\in\F_2^4:\quad
F(T_{(I,J,K,M)}) 
= {\ts{1\over2}} \sum_{i,j\in\F_2} (-1)^{iI+jJ}T_{(i,j,K,M)}.
$$
Similar formulas for $N\in\{3,4,6\}$ are given in 
\cite[(29)]{nawe01}.
\end{prop}
Comparison with the geometric mirror map in Prop.\ \ref{geoms}
now allows to explicitly relate twisted ground states to the generators 
$E_s^{(l)}$ of $\Pi_N\otimes\Q\,$:
\begin{prop}\label{c}\cite{nawe01}
As before, let $\gamma_{MS},\; F$ denote the action of mirror symmetry
as induced by fiberwise T-duality on the Kummer type lattice
$\Pi_N$ and the space of twisted ground states $\TT$, respectively.
The $\C$ linear map
$$
C: \Pi_N\otimes\C\rightarrow\TT, \quad
\forall\, s\in\SS,\; l\in\{1,\dots,n_s-1\}: \quad
C(E_s^{(l)}) = 
{\ts{1\over\sqrt{n_s}}} \sum_{k=1}^{n_s-1} \zeta_{n_s}^{lk} T_s^k
$$
obeys
$
FC=C\gamma_{MS}.
$
Moreover, if on cohomology we use the scalar product induced by the 
intersection form and on the Hilbert space
of the orbifold CFT we use the standard scalar product with \req{norm},
then  $C$ induces an anti-isometry.
The image of $\Pi_N$ under $C$ is invariant under Hermitean
conjugation.
\end{prop}
To understand the map $C$ it is useful to study the ``quantum'' $\Z_N$ symmetry
of the orbifold CFT. It has generator $\vartheta$ which acts
trivially on the untwisted sector and by 
$\vartheta T_s^l = \zeta_{n_s}^l T_s^l$ on twisted ground states. Performing
the $\Z_N$ orbifold construction by this ``quantum'' symmetry on the orbifold
CFT reproduces the original toroidal SCFT. Since we want to study geometric
features of our orbifold CFT we write out the induced $\vartheta$ action on
$\wh\Pi_N$ instead of $\Pi_N$:
$$
\forall\, s\in\SS:\quad\quad
\vartheta\left(\widehat{E}_s^{(l)}\right) = \left\{ 
\begin{array}{ll}
\widehat{E}_s^{(l+1)} & \mbox{ if } l<n_s-1,\\
\widehat\upsilon
-\sum\limits_{j=1}^{n_s-1}\widehat{E}_s^{(j)} & \mbox{ if } l=n_s-1.
\end{array}\right.
$$
In other words,
the ``quantum'' $\Z_N$  symmetry of a $\Z_N$ orbifold CFT has its geometric 
counterpart in the $\Z_{n_s}$ symmetry of the extended Dynkin diagram
$\wh A_{n_s-1}$ for the irreducible components of the exceptional divisor over
each $\Z_{n_s}$ type singularity.

This observation has a simple explanation\footnote{We thank David E. 
Berenstein for this comment.} in terms of the group algebra of 
$\Z_N$, if we restrict to the local picture over each $s\in\SS$.
First, introduce an auxiliary state $T_s^0$ which is
subject to \req{norm} (and therefore is \textsc{not} the vacuum state). Then 
isometrically extend $C$ by $C(\wh\ups):=\sqrt{n_s}\, T_s^0$ to obtain
\begin{eqnarray*}
\forall\,l\in\{1,\dots,n_s-1\}:\quad\quad\quad 
C(\wh E_s^{(l)}) &=& 
{\ts{1\over\sqrt{n_s}}} \sum_{k=0}^{n_s-1} \zeta_{n_s}^{lk} T_s^k,\\
C(\wh\ups-\sum_{j=1}^{n_s-1}\wh E_s^{(j)}) &=& 
{\ts{1\over\sqrt{n_s}}} \sum_{k=0}^{n_s-1} T_s^k.
\end{eqnarray*}
The extended map has the form of a discrete Fourier transform on
$\Z_{n_s}$, if it is interpreted as map on the group algebra $\C\Z_{n_s}$ of 
$\Z_{n_s}$ with
$T_s^l$ corresponding, say, to the conjugacy class of
$\zeta_{n_s}^l$. Then up to normalization the $n_s$ elements of $\C\Z_{n_s}$ 
listed above are just the idempotents in $\C\Z_{n_s}$,
which are known to be related
to the conjugacy classes by a discrete Fourier transform.

Next, let us investigate the action of $\Lambda^*\oplus\Lambda$
on cohomology which is induced by the Weyl algebra representation
\req{weyl} under the map $C$. It proves useful to work with a dual basis
$\varepsilon_s^{(l)}$ with respect to $\{\wh E_s^{(l)},\;l\in\{1,\dots,n_s-1\};
\;\wh\ups-\sum_{j=1}^{n_s-1}  \wh E_s^{(j)}\}$: 
As before, denote by 
$\{(\wh E_s^{(l)})^*\}\subset\wh\EE_s\otimes\Q$ 
the dual basis with respect to the fundamental system
$\{\wh E_s^{(l)}\}$ of $\wh\EE_s$, then
$$
\varepsilon_s^{(l)}:=\left\{
\begin{array}{ll}
\wups & \mbox{if } l=0,\\
\wups + (\wh E_s^{(l)})^* & \mbox{if } 1\leq l < n_s,
\end{array}\right\},
\quad\forall\, l\in\Z:\, \varepsilon_s^{(l+n_s)}=\varepsilon_s^{(l)}.
$$
The induced action is  given by
\begin{eqnarray}\label{bundleweyl}
\forall \, (\mu,\lambda) \in \Lambda^*\oplus\Lambda,
&&\!\!\!\!\!\!\!\!\!
\forall\, s\in\SS,\; l\in\{0,\dots,n_s-1\}: \quad\nonumber\\
W(\mu,\lambda) \varepsilon_s^{(l)} &=& \varepsilon_{s^\prime}^{(\mu(n_s s)+l)},
\quad 
s^\prime = s+(\id-\theta)^{-1}\lambda  
= s-{\ts{1\over n_s}} \sum_{k=1}^{n_s-1} k\theta^k\lambda.
\end{eqnarray}
Similarly to our explanation of Prop.\ \ref{z4emb}, there is a
straightforward interpretation 
\cite{nawe01} in terms of the classical McKay correspondence.
Indeed, each of the vectors $\varepsilon_s^{(l)}$ is the
$H^0(X,\Z)\oplus H^2(X,\Q)$ part of the Mukai vector for 
one of the locally free sheaves
near $s\in\SS$ that were constructed in \cite{gsve83}. More precisely,
identify $s\in\SS$ with the origin in $\C^2$ and consider
$Y=\wt{\C^2/\Z_{n_s}}$. By \cite{gsve83,kn85}, for each $\Z_{n_s}$ equivariant
flat line bundle on $\C^2$ there is a corresponding locally free sheaf
on $Y$. We choose a fixed 
generator of $\Z_{n_s}$ and assume 
that on the fiber of the bundle over $s$ it acts by
$\chi_s^{(l)}(z)=\zeta_{n_s}^l z$. Then $(\wh E_s^{(l)})^*$ is the
first Chern class of the associated bundle on $Y$, and 
$\varepsilon_s^{(l)}$ is the part of its Mukai vector relevant to our 
discussion.

We can now interpret the Weyl algebra action \req{bundleweyl} as an action
on line bundles over $X$ \cite{nawe01}. Indeed, it is the natural action of 
$\Lambda^*\oplus\Lambda$, where $\Lambda$ acts by translations by elements
of the torsion subgroup
of the Jacobian torus of $T$
on the underlying torus $T$, and $\Lambda^*$ acts by tensorizing
with fixed line bundles. Moreover, the indeterminacy in our formula for 
mirror symmetry in Prop.\ \ref{geoms} translates directly into the freedom
of choice of an origin in the affine space which parameterizes $\Z_N$
equivariant flat line bundles on the four torus \cite{nawe01}.

Summarizing, in Prop.\ \ref{c}
we have established the explicit map between geometric and 
conformal field theoretic data which characterize orbifolds:
In the local picture over each singularity $s\in\SS$, 
we have a correspondence between twist fields
$T_s^1,\dots,T_s^{n_s-1}$ and the vectors 
$\varepsilon_s^{(1)},\dots,\varepsilon_s^{(n_s-1)}$ 
which induces a
correspondence between non-trivial conjugacy classes in $\Z_{n_s}$ and 
a basis of $H^2(Y,\Q)$, $Y=\wt{\C^2/\Z_{n_s}}$. 
Since we have worked with fixed choices for roots
of unity this is just a realization of the ``dual'' McKay correspondence
proven more generally
by Ito and Reid \cite{itre94}: For $R\in\N$ let $\mu_R$ denote the
group of complex $R^{th}$ roots of unity. Then for each
finite subgroup $G$ of
$SL(n,\C)$, there is a $1:1$ correspondence
between junior conjugacy classes in $Hom(\mu_{|G|},G)$ and 
a basis of $H^2(Y,\Q)$, where $Y$ is a minimal model of $\C^n/G$.
Note that for our $\Z_{n_s}\subset SL(2,\C)$
all non-trivial conjugacy classes in $Hom(\mu_{n_s},\Z_{n_s})$ are junior.

We have been mostly working in the compact setting where
$X=\wt{T/\Z_N}$ with $T=\R^4/\Lambda$. Here, we found that our CFT
version of the dual McKay correspondence is 
compatible with the natural Weyl
algebra representations of $\Lambda^*\oplus\Lambda$ on twisted ground
states and $\Z_N$ equivariant flat line bundles on $T$, respectively.
\section{An instructive example}\label{example}
In Sect.\ \ref{mirror} we have given an overview of various approaches
to mirror symmetry on $K3$: Greene/Plesser's construction (GP) 
for Gepner type models, Aspinwall/Morrison's or Dolgachev's approach
(AMD) for $M$ polarized families with Rohsiepe's improvement for 
Calabi-Yau hypersurfaces in toric varieties, and Va\-fa/Wit\-ten's 
SYZ like approach (VW) which  we have applied to $\Z_N$ orbifold CFTs
in $\MM^{K3}$ \cite{nawe01}, as explained
in Sect.\ \ref{cftvsgeo}. Not only does each of these approaches refer to a
fairly different setting, but also is a comparison almost impossible
since e.g.\ the GP construction as such does not involve the choice 
of a specific geometric interpretation. In \cite{asmo94,asmo,as96} Aspinwall
and Morrison noted that a comparison does make sense 
for models that are invariant
under mirror symmetry; first, they should be such under all applicable
versions of mirror symmetry, and second one can directly compare the
induced map on the fields of the relevant SCFTs in a chosen
``reference geometric interpretation''. 

Aspinwall and Morrison 
use the Gepner model $(1)(5)(40)$, which has trivial GP group,
i.e.\ is GP mirror self dual. A translation into the AMD approach is
possible \cite{asmo,as96}, 
which should prove GP=AMD \cite{asmo} and thereby provide an 
element of $O^+(H^{even}(X,\Z))$ that is needed for the proof of 
Th.\ \ref{globmod}. Unfortunately, for this Gepner model there exists
no geometric  interpretation in terms of a $\Z_N$ orbifold CFT of some 
toroidal theory. 
We will therefore work with a different model:
Consider the Gepner model $(2)^4$; the GP group to orbifold by in order to
produce the mirror is $\Z_4^2$. The elements of order two in $\Z_4^2$ 
generate a group $H=\Z^2_2$; the orbifold of $(2)^4$ by this group $H$ is the
Gepner type model $(\wt 2)^4$. By 
\cite[Th.\ 3.7]{nawe00} it possesses a geometric interpretation as 
$\Z_2$ orbifold CFT, see Prop.\ \ref{vwself} below.
\begin{prop}\label{msself}
The Greene/Plesser group $H^\ast$ for $(\wt 2)^4=(2)^4/H$ agrees with 
$H=\Z_2^2$, i.e. $(2)^4/H^\ast=(2)^4/H$ is the
mirror for the Gepner type model $(\wt 2)^4$. 
In other words, $(\wt 2)^4$ is GP
mirror self dual.
\end{prop}
\begin{pr}
Recall the Greene/Plesser construction \cite{grpl90}. For a Gepner 
model $\CC=(k_1)\cdots(k_r)$ with central charge $3d,\; d\in\N\,$,
let $z_i$ denote the generator of the
$\Z_{k_i+2}$ phase symmetry for its $i^{th}$ minimal model factor.
Recall that $\prod_i z_i$ acts trivially on $\CC$, and that those
$\prod_i z_i^{a_i}$ with  $a_i\in\N\,,\; \prod_i \zeta_{k_i+2}^{a_i}=1$ are
called \textsc{algebraic}.
The mirror model of $\CC/H$ with algebraic $H$
then is $\CC/H^*$, where $H^*$ contains all algebraic $\prod_i
z_i^{b_i}$ which for all
$\prod_i z_i^{a_i}\in H$ obey $\prod_i \zeta_{k_i+2}^{a_ib_i}=1$. 

In our case with $\CC=(2)^4$ one checks that both $H$ and $H^*$ are
generated by $z_1^2z_2^2$ and $z_1^2z_3^2$, proving the assertion.
\end{pr}
It follows that $(\wt 2)^4$ is just as good a model to study mirror
symmetry on as $(1)(5)(40)$. Let us check with the AMD approach:
\begin{prop}\label{torself}
Let $\Delta\subset\R^3$ 
denote the reflexive polyhedron which is associated to 
$(\wt 2)^4$. Then $\Delta$ is related to its dual $\Delta^*$ by a
$GL(3,\Z)$ transformation. In other words, $(\wt 2)^4$ is AMD mirror self
dual.

The corresponding families of toric Calabi-Yau hypersurfaces possess
generic (toric) Picard lattice\footnote{For\label{hyperbolic} 
$n\in\Z$ the lattice of rank $1$
generated by a vector of length squared $n$ is denoted
$\langle n \rangle:=\Z(n)$. Moreover,
$\DD_{2k}$ denotes the lattice $\DD_{2k}:=\{x\in\Z^{2k}\mid\sum_i x_i\equiv
0(2)\}$, and $U$ is
the \textsc{hyperbolic lattice} generated by two null vectors with
scalar product one.}
\begin{equation}\label{picd}
\begin{array}{rclclcl}
Pic_{tor}(\Delta) &\!\!\!=\!\!\!& \check{Pic}(\Delta^*)
&\!\!\!=\!\!\!&\hphantom{-}\langle4\rangle\oplus\langle-2\rangle^6
&\!\!\!\cong\!\!\!&Pic_{tor}(\Delta^*),\\[3pt]
Pic(\Delta) &\!\!\!=\!\!\!& \check{Pic}_{tor}(\Delta^*)
&\!\!\!=\!\!\!&\langle-4\rangle\oplus\langle-2\rangle^4\oplus
\DD_6(-1)\oplus U 
&\!\!\!\cong\!\!\!&Pic(\Delta^*).
\end{array}
\end{equation}
\end{prop}
\begin{pr}
The Abelian group action associated to the mirror of
$(2)^4$ is the diagonal $\C^*$ action on $\C^4$. 
Additionally,  for $(\wt 2)^4$ there
is a $\Z_2^2$ action which is
specified by the generators $z_1^2z_2^2$ and $z_1^2z_3^2$ that we
used in the proof of Prop.\ \ref{msself}, where $z_i$ now acts 
by multiplication with $\zeta_{k_i+2}$ on the $i^{th}$ component of
$\C^4$. This induces three independent relations on $\C^4$, and in
$\Z^3\subset\R^3$ yields $\Delta^*$ as  convex hull of
$$
\left(\begin{array}{c}1\\0\\0\end{array}\right),
\left(\begin{array}{c}1\\2\\0\end{array}\right),
\left(\begin{array}{c}1\\0\\2\end{array}\right),
\left(\begin{array}{c}-3\\-2\\-2\end{array}\right)
\quad\in\quad\R^3.
$$
It is a regular tetrahedron with edges of length $2$ in lattice
units. One finds
\begin{eqnarray}\label{deltastar}
\hphantom{\Delta\Delta\Delta\Delta}
\Delta &=& \mbox{conv. hull} \left\{
\left(\begin{array}{c}3\\-2\\-2\end{array}\right),
\left(\begin{array}{c}-1\\2\\0\end{array}\right),
\left(\begin{array}{c}-1\\0\\2\end{array}\right),
\left(\begin{array}{c}-1\\0\\0\end{array}\right)
\right\} = A\Delta^*,\\
A&=&
\left(\begin{array}{ccc}
3&-2&-2\\ -2&2&1\\ -2&1&2 
\end{array}\right)  \in GL(3,\Z),\nonumber
\end{eqnarray}
proving that $(\wt 2)^4$ is AMD mirror self dual.

\req{picd} follows from Prop.\ \ref{falksms} together with
the methods of \cite{ba94,pesk97,ro01b}. Namely, by Prop.\ \ref{falksms}
and the above,
$$
Pic(\Delta^*)\cong Pic(\Delta)=\check{Pic}_{tor}(\Delta^*),
\quad
{Pic}_{tor}(\Delta^*)\cong {Pic}_{tor}(\Delta)
=\check{Pic}(\Delta^*),
$$
and $Pic_{tor}(\Delta)$ can be directly read off from the lattice vectors on 
edges of $\Delta$.
First, three appropriate lattice points have to be removed\footnote{We
always  omit the origin $0\in\Delta$ as well.}, since they
correspond to mere coordinate transformations. Here, we can omit one 
vertex and two neighboring centers of edges. The remaining diagram provides a
Dynkin type diagram for the generators of $Pic_{tor}(\Delta)$. With
the techniques explained in \cite{ba94,pesk97,ro01b} one
checks that each vertex corresponds to a toric divisor of self-intersection
number $-2$, whereas centers of edges correspond to toric divisors
with self-intersection number $-4$. Each connecting line in our diagram
corresponds to an intersection number $2$. One then checks that the
resulting lattice is $Pic_{tor}(\Delta)$ as given in \req{picd}.

By the results of \cite{ba94,pesk97,ro01b} each
lattice vector which is an inner point in an edge $\theta$ of
$\Delta$ will correspond to a toric divisor on our $K3$ surface which splits
into $k$ disjoint divisors, where $k$ is the length (in lattice units)
of the edge
$\theta^*$ of $\Delta^*$ which is dual to $\theta$. If $k>1$, all
intersection numbers are divided by $k$ to obtain the intersection
numbers for these non-toric components. Therefore in our case each inner
point of an edge corresponds to a toric divisor which
splits into two disjoint non-toric ones with self-intersection number
$-2$ each. With a somewhat lengthy calculation, 
the resulting lattice is checked to agree with
$Pic(\Delta)$ as given in \req{picd}.  
\end{pr}
To compare with the VW approach to mirror symmetry we need to 
have an appropriate geometric interpretation of $(\wt 2)^4$ as
$\Z_N$ orbifold CFT of some toroidal theory. Indeed,
\begin{prop}\label{vwself}
The Gepner type model $(\wt 2)^4$ admits a geometric interpretation
as $\Z_2$ orbifold CFT of the toroidal model 
on $T=\R^4/\Lambda,\; \Lambda={1\over\sqrt2}\DD_{4}$ with the B-field
$B^*$ given in \req{bstar} for which the theory has enhanced symmetry by the 
Frenkel-Kac
mechanism.
It is  VW  mirror self dual.
\end{prop}
\begin{pr}
Together with W.~Nahm we have proven the first part of the proposition 
in \cite[Th.\ 3.7]{nawe00}. 

As to the second part, we first have to modify our 
construction in Sects.\ \ref{mirror},\ \ref{cftvsgeo}, since the
underlying  toroidal model has non-vanishing B-field and
non-orthogonal lattice $\Lambda={1\over\sqrt2}\DD_{4}$. However, 
${\sqrt2}\Z^4\subset{1\over\sqrt2}\DD_4\subset{1\over\sqrt2}\Z^4$.
Therefore, we can consider the mirror map induced by
fiberwise T-duality on the first two
coordinates
of the  orthogonal torus $\R^4/{1\over\sqrt2}\Z^4$. Since 
with respect to the standard basis of $\R^4$
\begin{equation}\label{bstar}
B^*=\left(\begin{array}{cccc}
0&1&0&0\\-1&0&0&0\\0&0&0&1\\0&0&-1&0
\end{array}\right): \;
\Lambda\otimes\R\longrightarrow\Lambda^*\otimes\R\,,
\end{equation}
this mirror map 
corresponds to T-duality on a toroidal SCFT in two real dimensions
with lattice ${1\over\sqrt2}\DD_{2}$ and B-field 
$B^{**}=\left(\begin{array}{cc}
0&1\\-1&0\end{array}\right)$. The latter model again has enhanced 
symmetry due to the Frenkel-Kac mechanism and is invariant under
T-duality. All formulas derived for orthogonal tori with vanishing
B-field can therefore be applied, and the model is invariant under
this version of the VW approach to mirror symmetry on $K3$.
\end{pr}
\begin{rem}
As remarked in \cite[Rem.\ 3.8]{nawe00}, the bosonic
subtheory of $(\wt 2)^4$ agrees with the bosonic subtheory of the
toroidal model on $\R^4/\Z^4$ with vanishing B-field. The latter
toroidal SCFT  is also  VW mirror self dual.
\end{rem}
To summarize, the Gepner type model $(\wt 2)^4$ allows a discussion
of all three versions of mirror symmetry on $K3$ that where mentioned
in Sect.\ \ref{mirror}. It is mirror self dual under all of them. 

Now let us
compare the induced maps on the fields of the theory:
\begin{prop}\label{compms}
On $(\wt 2)^4$, the mirror maps induced by the VW and GP approaches
to mirror symmetry agree. 
\end{prop}
\begin{pr}
Recall \cite[Th.\ 3.7]{nawe00} that $(\wt 2)^4$
possesses a left and a right current algebra of type $su(2)_1^6$, 
such that (up to normalizations)
each field in the theory is entirely specified by its
charges under a left and a right $u(1)^6$ current algebra. 
Moreover, mirror symmetry only acts on the left hand side of each
field. Hence it suffices to determine the map induced by mirror symmetry on 
the left handed current algebra
$su(2)_1^6$.

As in the proof of \cite[Th.\ 3.7]{nawe00} we denote the $U(1)$
currents in the left handed $su(2)_1^6$ by $J,\;A,\;P,\;Q,\;R,\;S$, and 
the other two generators of each $su(2)_1$ factor by 
$J^\pm,\;A^\pm,\;P^\pm,\;Q^\pm,\;R^\pm,\;S^\pm$.

Let us begin with the Gepner type description of $(\wt 2)^4$ and the
GP approach to mirror symmetry. Let $J_1,\dots,J_4$ denote the $U(1)$
currents of the minimal model factors in $(\wt 2)^4$. 
There are primary fields $\Phi^l_{m,s;\qu m,\qu s}$ with quantum numbers
$l\in\{0,1,2\},\; m,\qu m\in\Z_8,\;s,\qu s\in\Z_4$ 
in each minimal model factor. 
Moreover,
$X_{i,j}$ denotes the primary field with factor $\Phi^0_{4,2;0,0}$ in the
$i^{th}$ and $j^{th}$ component and $\Phi^0_{0,0;0,0}$ elsewhere,
and for $Y_{i,j}$ we have $\Phi^0_{-2,2;0,0}$ in the
$i^{th}$ and $j^{th}$ factor and $\Phi^0_{2,2;0,0}$ otherwise.
The formula between \cite[(3.18)]{nawe00} and \cite[(3.19)]{nawe00} 
gives the $su(2)_1^6$ current algebra in terms of the $J_i,\;
X_{i,j},\;Y_{i,j}$.
Now note that the GP version of mirror symmetry is induced by 
$J_i\mapsto -J_i,\; i\in\{1,\dots,4\}$, and therefore 
$Y_{i,j} \leftrightarrow Y_{k,m}$ with $\{i,j,k,m\}=\{1,\dots,4\}$,
whereas the $X_{i,j}$ are invariant. Hence, the induced map on
$su(2)_1^6$ is given by 
\begin{eqnarray}\label{msex}
(\JJ,\JJ^+,\JJ^-)&\longmapsto& (-\JJ,\JJ^-,\JJ^+)
\mbox{ for $\JJ\in\{J,A,P,Q\}$,}\\
\mbox{and }\quad
(\JJ,\JJ^+,\JJ^-)&\longmapsto& \hphantom{-}(\JJ,\JJ^+,\JJ^-)
\mbox{ for $\JJ\in\{R,S\}$.}\nonumber
\end{eqnarray}
In the nonlinear sigma model description of $(\wt 2)^4$, the
$su(2)_1^6$ current algebra is obtained from $\Z_2$ invariant fields
in the underlying toroidal SCFT $\CC_T$. 
There, we have four Majorana fermions
$\psi^1, \dots, \psi^4$ and their superpartners, four Abelian
$U(1)$ currents $j^1,\dots,j^4$. Let $e_1,\dots,e_4$ denote the standard 
orthonormal generators of $\Z^4$. 
As before, we identify $H^1(T,\Z)\cong\Lambda^*$, and generators of
$\Lambda^*$ are expressed in terms of the $e_i$, such that we can interpret the
$e_i$ as elements of $H^1(T,\R)$.
Recall \cite{hala82} that our
torus fibration with fiber direction $e_1,e_2$
is special Lagrangian with respect to the complex structure
$\II$ given by $(e_1-ie_3)\wedge(e_2+ie_4)$ (see  \cite{nawe01}). 
Accordingly, we define
$$
\psi_\pm^{(1)}:= {\ts{1\over\sqrt2}} (\psi^3\pm i\psi^1), \quad
\psi_\pm^{(2)}:= {\ts{1\over\sqrt2}} (\psi^2\pm i\psi^4).
$$
Hence in comparison to \cite[(2.1)]{nawe00} we have to rename
\begin{equation}\label{rename}
(\psi^1,\psi^2,\psi^3,\psi^4) \mapsto (\psi^3,\psi^1,\psi^2,\psi^4),
\quad
(j^1,j^2,j^3,j^4) \mapsto (j^3,j^1,j^2,j^4)  .
\end{equation}
The $\Z_2$ invariant current algebra $su(2)_1^2$ obtained from the
fermions is given in \cite[(2.2)]{nawe00}.

Vertex operators in the toroidal
SCFT $\CC_T$
are specified by their charges with respect to $(j^1,\dots,j^4)$. Set
$$
\forall\;i,j\in\{1,\dots,4\},\;i\neq j:\quad
\alpha_{i,j}^\pm := \sqrt2 (e_i\pm e_j).
$$
Then the $24$ holomorphic vertex operators with left dimension $1$
in $\CC_T$ have left handed
charges $\alpha_{i,j}^\pm,\; -\alpha_{i,j}^\pm$ with respect to 
$(j^1,\dots,j^4)$. The $\Z_2$ invariant vertex operator with charges
$\alpha_{i,j}^\pm,\; -\alpha_{i,j}^\pm$ is denoted $U_{i,j}^\pm$,
and $W_{i,j}^\pm:=U_{i,j}^+\pm U_{i,j}^-$ after appropriate normalization. With
\cite[(3.18)]{nawe00} 
one finds the representation of the missing
$su(2)_1^4$ in terms of the $U_{i,j}^\pm$. To match with our orbifold
conventions, we still have to take \req{rename} into account and perform a 
permutation of the factors of $su(2)_1^4$. Moreover, in the proof of
Prop.\ \ref{amdms} we will need a specific choice of the Cartan torus in each
$su(2)_1$ factor: The choice of complex structure $\II$ used in the 
VW approach to mirror symmetry must match with the choice of
$N=2$ superconformal algebra in the $N=4$ superconformal algebra that
is obtained from the Gepner construction.
This leaves us with a renaming of the $su(2)_1^4$ generators in
$\CC_T/\Z_2$ by
\begin{eqnarray*}
(P,\hphantom{^\pm}Q,\hphantom{^\pm}R,\hphantom{^\pm}S)
&\!\!\!\!\mapsto\!\!\!\!& \hphantom{^\pm}(R,\hphantom{S\mp}\;
{\ts{i\over2}}(S^+-S^-),\hphantom{P\mp}\;
{\ts{i\over2}}(P^+-P^-),\hphantom{Q\mp}\;{\ts{1\over2}}(Q^++Q^-)); \\
(P^\pm,Q^\pm,R^\pm,S^\pm)&\!\!\!\!\mapsto\!\!\!\! &
(R^\pm,S\mp{\ts{i\over2}}(S^++S^-),P\mp{\ts{i\over2}}(P^++P^-),
Q\mp{\ts{1\over2}}(Q^+-Q^-))
\end{eqnarray*}
with respect to \cite[(3.18)]{nawe00}. Now we can write out the 
appropriate expressions for all $\JJ,\;\JJ^\pm$ in \req{msex} in terms
of fields in the nonlinear sigma model.

From our explanation in the proof of Prop.\ \ref{vwself}, the VW mirror
map is induced by $(j^1,j^2,j^3,j^4)\mapsto (-j^1,-j^2,j^3,j^4)$,
$(\psi^1,\psi^2,\psi^3,\psi^4)\mapsto(-\psi^1,-\psi^2,\psi^3,\psi^4)$.
It therefore leaves $U_{1,2}^\pm$ and $U_{3,4}^\pm$ invariant and
on all other $U_{i,j}^\pm$ induces $U_{i,j}^+\leftrightarrow U_{i,j}^-$.
One now checks that the induced map on the $su(2)_1^6$ current algebra
in fact agrees with \req{msex}, 
proving the assertion.
\end{pr}

\begin{prop}\label{amdms}
The AMD approach  to mirror symmetry on $(\wt 2)^4$ is
compatible with the GP$\stackrel{\mbox{\scriptsize Prop.\ \ref{compms}}}{=}$VW 
approaches.
\end{prop}
\begin{pr}
To study the AMD approach to mirror symmetry on $(\wt 2)^4$
we invoke the toric
description given in Prop.\ \ref{torself}. In the following, we will use the
notations introduced in Props.\ \ref{torself} and \ref{compms}. Let us denote
the four vertices of $\Delta$ in the order they are listed in \req{deltastar}
by $A,\;B,\;C,\;D$, and the center of the edge
between $A$ and $B$ by $AB$ etc. The origin $0\in\Delta$ is denoted
$o$. The corresponding elements of $Pic_{tor}(\Delta)$ are
$\omega_A,\;\omega_B,\dots; \omega_{AB},\; \omega_{AC}, \dots; \omega_o$.
By the proof of Prop.\ \ref{torself}
$$
\begin{array}{rclcl}
\omega_A^2 &=& \omega_B^2 \quad=\quad \omega_C^2 \quad=\quad \omega_D^2   
&=& -2, \\[3pt]
\omega_{AB}^2 
&=& \omega_{AC}^2 \;=\; \omega_{AD}^2 \;=\; \omega_{BC}^2 \;=\;\omega_{BD}^2 
\;=\;\omega_{CD}^2
&=& -4, \\[3pt]
\omega_o^2 
&=& (\omega_A +\cdots+ \omega_D + \omega_{AB} +\cdots+ \omega_{CD})^2 
&=& 16.
\end{array}
$$
On inspection of the relations between the various cocycles $\omega_\bullet$
one finds
$$
\omega_o
= -2 (-2\omega_A+2\omega_B+2\omega_C 
- \omega_{AD} + 2\omega_{BC} + \omega_{BD} + \omega_{CD}) 
= 2 \omega_{\wt o}
$$
with $\omega_{\wt o}\in Pic_{tor}(\Delta^*)$.
Moreover,
\begin{eqnarray}\label{decompose}
\omega_A 
&=& -{\ts{1\over2}} ( \omega_{\wt o}+\omega_{AB}+\omega_{AC}+\omega_{AD} ),
\nonumber\\
\omega_B
&=& -{\ts{1\over2}} ( \omega_{\wt o}+\omega_{AB}+\omega_{BC}+\omega_{BD} ),\\
\omega_C
&=& -{\ts{1\over2}} ( \omega_{\wt o}+\omega_{CD}+\omega_{AC}+\omega_{BC}  ),
\nonumber\\
\omega_D
&=& -{\ts{1\over2}} ( \omega_{\wt o}+\omega_{CD}+\omega_{AD}+\omega_{BD}  ).
\nonumber
\end{eqnarray}
The AMD approach to mirror symmetry in its improved version by 
Rohsiepe (see Prop.\ \ref{falksms})  implies 
that the mirror map induces a lattice 
automorphism which acts on
$$
Pic_{tor}(\Delta)\perp Pic(\Delta^*)\subset
\Gamma^{2,18}\subset H^2(X,\Z)
$$
such that $Pic_{tor}(\Delta)$ is mapped onto the sublattice
$Pic_{tor}(\Delta^*)\subset Pic(\Delta^*)$ of rank $7$.

To compare with the GP approach to mirror symmetry note that the mirror of
$(2)^4$ corresponds to a family of Calabi-Yau toric hypersurfaces with
generic Picard lattice of rank $1$. In $(2)^4$ we therefore have 
$20-1=19$ states with conformal dimensions $(1/4,1/4)$ which are
uncharged  under
the $U(1)$ current of the $N=2$ superconformal algebra 
specified by the Gepner construction and which also belong to
$(2)^{\otimes4}$. As we shall explain below, by the correspondence
between Gepner states and complex structure deformations for the mirror 
in terms of
monomials \cite{ge87,ge88} together with
the monomial-divisor map \cite{agm93},
these $19$ states correspond to 
generators of the 
\textsc{quantum Picard lattice} $Pic\oplus U$ associated to $(2)^4$
(see \cite{as96} and recall from footnote \ref{hyperbolic} that
$U$ denotes the \textsc{hyperbolic lattice} which is
generated by two null vectors with scalar product $1$).
One of these $19$ states 
generates deformations of the volume of the relevant toric 
Calabi-Yau hypersurface and does not correspond to a vector in the ordinary
Picard lattice $Pic$. To generate the latter, an appropriate element of the
K\"ahler cone has to be determined separately. 

The family corresponding
to $(\wt 2)^4$ is obtained from that corresponding to $(2)^4$
by a $\Z_2^2$ orbifold procedure and has generic Picard
lattice $Pic(\Delta)$  of rank $13$ as in \req{picd}.  
Among the $19$ states mentioned before,
seven are invariant under the $\Z_2^2$ action and
therefore should correspond to elements of 
$Pic_{tor}(\Delta)\oplus U = \check{Pic}(\Delta^*)\oplus U$
above: Six of type
$\Xi_{i,j}$ with $\Phi^0_{1,1;1,1}$ in the $i^{th}$ and $j^{th}$
factors and $\Phi^0_{-1,-1;-1,-1}$ elsewhere, and 
$(\Phi^1_{2,1;2,1})^{\otimes4}$.
The geometric counterpart 
of each of these Gepner states in the family described by $\Delta$
is determined by the 
Abelian symmetries of $(\wt 2)^4$. More precisely, we can apply
the operator $(\Phi^0_{-1,-1;-1,-1})^{\otimes4}$ of spectral flow to obtain
states in the $(c,c)$ ring of the
Gepner type model and compare with deformations of the
complex structure for members of the mirror family. 
By the
monomial-divisor map \cite{agm93}
they correspond to  specific elements of 
$Pic_{tor}(\Delta)\oplus U$,
and if these elements belong to $Pic_{tor}(\Delta)$ they
can be labeled by lattice vectors in $\partial\Delta$.
Here we have $\wt\Xi_{i,j}$ with 
$\Phi^0_{-2,-2;-2,-2}\sim\Phi^2_{2,0;2,0}$ in the 
$i^{th}$ and $j^{th}$
factors and $\Phi^0_{0,0;0,0}$ elsewhere which 
by \cite{ge87,ge88} corresponds to 
the monomial $x_i^2x_j^2\in\C[x_A,x_B,x_C,x_D]$,
or, in terms of $\Delta$, the center of an edge.
The remaining state is mapped to $(\Phi^1_{1,0;1,0})^{\otimes4}$ 
under spectral flow and
corresponds to the generator of volume deformations of the toric 
Calabi-Yau hypersurface which will not be discussed in the following.

Recall from the proof of Prop.\ \ref{torself}
that for a  choice of generators of $Pic_{tor}(\Delta)$ at most
one vertex of $\Delta$ can be omitted
such that the six lattice vectors corresponding to the $\Xi_{i,j}$
cannot be expected to generate a primitive sublattice of
$Pic_{tor}(\Delta)$.
However, using the exact field-to-field identifications of
\cite[Th.\ 3.7]{nawe00} we can explicitly determine  
the geometric counterparts of 
these states in our nonlinear sigma model interpretation of $(\wt 2)^4$, 
which will enable us to calculate intersection numbers and
perform a compatibility check between the AMD and the 
GP$\stackrel{\mbox{\scriptsize Prop.\ \ref{compms}}}{=}$VW approaches to mirror
symmetry. 
Here, $(t_1,t_2,t_3,t_4)\in\F_2^4$
labels the fixed point ${1\over2}\sum_j t_j\lambda_j$ with appropriate
generators $\lambda_j$ of the torus lattice
$\Lambda={1\over\sqrt2}\DD_4$
with $T=\R^4/\Lambda$: 
$$
\lambda_1 := {\ts{1\over\sqrt2}} ( e_1+e_3),\;
\lambda_2 := {\ts{1\over\sqrt2}} ( e_1-e_3),\;
\lambda_3 := {\ts{1\over\sqrt2}} ( e_2+e_3),\;
\lambda_4 := {\ts{1\over\sqrt2}} ( e_4-e_1).
$$
As before, by $e_1,\dots,e_4$ we denote the standard orthogonal
generators of $\Z^4\subset\R^4$, where $\R^4$ is identified with its
dual by the use of the standard scalar product.
By $\mu_1,\dots,\mu_4$ we denote the dual
basis with respect to $\lambda_1,\dots,\lambda_4$ which is readily
interpreted as basis of $H^1(T,\Z)$.
Then, four of the above mentioned six
states $\Xi_{i,j}$ are identified with linear combinations of
twist fields $T_t,\; t\in\F_2^4$, in the nonlinear sigma model
description of $(\wt 2)^4$. From the proof of \cite[Th.\ 3.7]{nawe00}
together with the renaming
that was observed in the proof of Prop.\ \ref{compms} we find the explicit 
linear combinations (see the formula below). From Prop.\ \ref{c} we know that
$T_t,\; t\in\F_2^4$, corresponds to a cocycle $(-E_t)\in H^{even}(X,\R)$ of
the relevant Kummer surface $X$. With \req{ehat} it follows that
$\wh E_t=E_t+{1\over2}\wh\ups\in H^2(X,\Z)$, and all in all we find
that the following identifications can be made in $H^2(X,\Z)$:
\begin{eqnarray*}
\Xi_{1,3}\quad\widehat{=}\quad\omega_{AC}
&=& 
{\ts{1\over2}}\sum_{j,k\in\F_2} (-1)^{j+k} \wh E_{(j,j,k,k)}
+ {\ts{1\over2}}\sum_{j,k\in\F_2} (-1)^{j+k} \wh E_{(j,j,k,k+1)},\nonumber\\
\Xi_{2,4}\quad\widehat{=}\quad\omega_{BD}
&=& 
{\ts{1\over2}}\sum_{j,k\in\F_2} (-1)^{j+k} \wh E_{(j,j,k,k)}
- {\ts{1\over2}}\sum_{j,k\in\F_2} (-1)^{j+k}\wh E_{(j,j,k,k+1)},\nonumber\\
\Xi_{1,4}\quad\widehat{=}\quad\omega_{AD}
&=& 
{\ts{1\over2}}\sum_{j,k\in\F_2} (-1)^{j} \wh E_{(j,j,k,k)}
+ {\ts{1\over2}}\sum_{j,k\in\F_2} (-1)^j \wh E_{(j,j,k,k+1)},\\
\Xi_{2,3}\quad\widehat{=}\quad\omega_{BC}
&=& 
{\ts{1\over2}}\sum_{j,k\in\F_2} (-1)^{j} \wh E_{(j,j,k,k)}
- {\ts{1\over2}}\sum_{j,k\in\F_2} (-1)^j \wh E_{(j,j,k,k+1)},\nonumber\\
\Xi_{1,2}\quad\widehat{=}\quad\omega_{AB}
&=&
\sqrt2\left(\; e_1\wedge(e_3-e_2) \;+\; (e_2+e_3)\wedge e_4\;\right)\\
&=&
\sqrt2\left(\;
(\mu_1-\mu_2)\wedge\mu_4 \;-\; \mu_1\wedge\mu_2 \;+\; \mu_3\wedge\mu_4
\;\right),\nonumber\\
\Xi_{3,4}\quad\widehat{=}\quad\omega_{CD}
&=&
\sqrt2\left(\; e_1\wedge(e_2+e_3) \;+\; (e_2-e_3)\wedge e_4\;\right)\\
&=&\sqrt2\left(\;
(\mu_1+\mu_2)\wedge\mu_3 \;-\; \mu_1\wedge\mu_2 \;+\; \mu_3\wedge\mu_4\;
\right),\nonumber\\
\omega_{\wt o} 
&=&
\sqrt2\left(\; e_1\wedge(e_2-e_3) \;+\; (e_2+e_3)\wedge e_4\;\right)\\
&=& \sqrt2\left(\; \mu_1\wedge\mu_2 \;+\; \mu_3\wedge\mu_4\;
\right).
\end{eqnarray*}
One first checks that the GP=VW approach to mirror symmetry indeed maps
any rank $7$ lattice $P\subset H^{even}(X,\Z)$
containing these seven vectors onto a lattice in $P^\perp$.
Moreover, all of them are pairwise orthogonal primitive lattice vectors, and 
$\omega_{\wt o}^2=4$, whereas for the other $\omega_\bullet$ listed above 
we have $\omega_\bullet^2=-4$, in accordance with the toric picture.
To show that the primitive sublattice of $H^2(X,\Z)$
containing these vectors indeed is 
$\langle4\rangle\oplus\langle-2\rangle^6\cong 
Pic_{tor}(\Delta)$, we need to show that the corresponding
$\omega_A,\dots,\omega_D$ given in \req{decompose} are lattice vectors. Indeed,
\begin{eqnarray*}
\omega_A 
&=& -\pi_\ast (\mu_3\wedge\mu_4) \\
&&\;
- {\ts{1\over2}}\pi_\ast([\mu_1-\mu_2]\wedge\mu_4)
-{\ts{1\over2}} \left( 
\wh E_{(0,0,0,0)}- \wh E_{(1,1,0,0)}+ \wh E_{(0,0,0,1)}- \wh E_{(1,1,0,1)} \right),\\
\omega_B
&=& -\pi_\ast (\mu_3\wedge\mu_4) \\
&&\;
- {\ts{1\over2}}\pi_\ast([\mu_1-\mu_2]\wedge\mu_4)-{\ts{1\over2}} \left( 
\wh E_{(0,0,0,0)}- \wh E_{(1,1,0,0)}- \wh E_{(0,0,0,1)}+ \wh E_{(1,1,0,1)} \right),\\
\omega_C
&=& -\pi_\ast (\mu_3\wedge\mu_4) \\
&&\;
- {\ts{1\over2}}\pi_\ast([\mu_1+\mu_2]\wedge\mu_3)-{\ts{1\over2}} \left( 
\wh E_{(0,0,0,0)}- \wh E_{(1,1,0,0)}- \wh E_{(0,0,1,0)}+ \wh E_{(1,1,1,0)} \right),\\
\omega_D
&=& -\pi_\ast (\mu_3\wedge\mu_4) \\
&&\;
- {\ts{1\over2}}\pi_\ast([\mu_1+\mu_2]\wedge\mu_3)-{\ts{1\over2}} \left( 
\wh E_{(0,0,0,0)}- \wh E_{(1,1,0,0)}+ \wh E_{(0,0,1,0)}
- \wh E_{(1,1,1,0)} \right)
\end{eqnarray*}
are of the type listed in \req{split}.
\end{pr}
\section{Conclusions}\label{conc}
In this note, we have given a brief description of the known structure
of the moduli space $\MM$ of $N=(4,4)$ SCFTs with $c=6$ and have spelled out
some of its explicit connections to geometry.

The algebraic structure of $\MM$ had been known before 
\cite{na86,se88,ce91,asmo94}, but only recently \cite{nawe00,we00} the 
location of orbifold CFTs on $K3$ that are obtained from toroidal theories
in $\MM$ was described in terms of geometric quantities. 
The main problem is the determination of the
B-field values in a geometric interpretation of such an orbifold CFT on
the corresponding orbifold limit $X$ of $K3$. The B-field is nonzero
in direction of each component of the exceptional divisor in $X$
\cite{as95} and can be determined explicitly
by a generalization of Nikulin's methods
to describe the Kummer lattice and its embedding in $H^2(X,\Z)$.
We  argue that these nonzero B-field values can be understood as
artifact from the specific choice of geometric interpretation on the 
orbifold limit $X$. Moreover, they have a straightforward explanation
in terms of the classical McKay correspondence which we venture to conjecture
should allow a determination of the B-field values in a more general setting,
too. This also provides an explicit geometric understanding of B-fields,
at least in the context of SCFTs on $K3$.

The second part of this note is devoted to a discussion of mirror symmetry
on the $\Z_N$ orbifold CFTs discussed before. We investigate a
version of mirror symmetry on elliptically fibred $K3$ surfaces that is 
induced by fiberwise T-duality on nonsingular fibers. This part of the note 
is a summary of
\cite{nawe01}. Our explicit knowledge of the relevant lattices allows the
determination of the corresponding automorphism of the lattice of integral
cohomology on $K3$. On the other hand, the action  
on the relevant states of our SCFTs is found to have the structure
of a fiberwise discrete $\Z_N$ type Fourier transform. 
A comparison of the
geometric and the conformal field theoretic mirror maps now provides us with
a dictionary to directly translate geometric data into conformal field 
theoretic ones. The ``quantum'' $\Z_N$ symmetry of the twisted sector
of $\Z_N$ orbifold CFTs is confirmed to have a geometric meaning. Moreover,
the action of the toroidal vertex operator algebra on twisted ground states
of the orbifold CFT translates into a natural action on line bundles on
$K3$ which are obtained by the classical McKay correspondence from 
$\Z_N$ equivariant line bundles on the underlying torus. In fact, our
dictionary can be interpreted as CFT version of Ito/Reid's ``dual'' McKay
correspondence \cite{itre94}. 
It bears the additional property of being compatible with the Weyl algebra
representation of the toroidal vertex operator algebra on twisted ground
states and $\Z_N$ equivariant flat line bundles, respectively. This
is also the explicit map that has to be
used in order to resolve the objection of \cite{fago01} to Ruan's 
conjecture \cite{ru00} on the orbifold cohomology of hyperk\"ahler 
surfaces\footnote{The fact that our transformation resolves this objection
was explained to us by Yongbin Ruan \cite{ru01} and goes back to an
earlier observation by Edward Witten.} (see \cite{nawe01}).

Apart from our approach to mirror symmetry, which is based on ideas by
Vafa and Witten \cite{vawi95}, the literature contains many statements about
mirror symmetry on $K3$ which at first sight appear not to be compatible.
Therefore, in this note we have included the discussion of an example
which allows for a comparison of our approach with two other mainstream
versions of mirror symmetry, due to Greene and Plesser \cite{grpl90} on the 
one  hand and Aspinwall/Morrison and Dolgachev \cite{asmo94,do96} on the other.
In fact, with an emendation of the latter approach 
in a toric setting \cite{ba94} due to Rohsiepe
\cite{ro01a,ro01b} we can show that all three versions of mirror symmetry
on $K3$ agree for our example, at least to the extent of comparability.

The virtue of our particular example is the fact that it is mirror self
dual. There are other examples of SCFTs in $\MM$ which allow the application
of all three versions of mirror symmetry but are not as well behaved
with respect to a comparison. E.g., the Gepner model $(2)^4$ has a nonlinear
sigma model interpretation as the $\Z_4$ orbifold CFT of the 
toroidal model on $\R^4/\Z^4$ with vanishing B-field \cite[Th.\ 3.5]{nawe00}.
This model appears to be mirror self dual under our approach to mirror 
symmetry since the underlying toroidal theory is, but it is not mirror 
self dual under the other two versions mentioned above. The resolution to
this puzzle probably
again is the special role of the B-field for orbifold CFTs on $K3$:
It is known that a shift of the B-field $B\in H^2(X,\R)$ in one of our
SCFTs by an integral two form does not change the physics in our theory.
Therefore, in \cite{nawe00,we00,nawe01} we have effectively considered the 
B-field as element of $H^2(X,\R)/H^2(X,\Z)$. 
In particular, we do not keep track
of integral B-field shifts which might occur under mirror symmetry on an
orbifold CFT of a toroidal theory
$\CC_T/\Z_N$, even if $\CC_T$ is mirror self dual. Since we
have given a geometric interpretation of the role of the B-field by using
the classical and the dual McKay correspondence, it might be interesting to
study this phenomenon in greater detail.

Summarizing, we hope to have convinced the reader that orbifold CFTs on 
$K3$ are simple enough to make explicit mathematical statements and
complicated enough to provide a rich playground for a study of 
interrelations between
geometry and conformal field theory.
%
%
\bibliographystyle{kw}
\bibliography{kw} 
\end{document}